\documentclass{article}
\usepackage{arxiv}
\usepackage[utf8]{inputenc} 
\usepackage[T1]{fontenc}    
\usepackage{hyperref}       
\usepackage{url}            
\usepackage{booktabs}       
\usepackage{amsfonts}       
\usepackage{nicefrac}       
\usepackage{microtype}      
\usepackage{setspace}
\usepackage{amsmath,amsthm}
\usepackage[linguistics]{forest}
\usepackage[mathscr]{euscript}
\usetikzlibrary{fit}
\usepackage{float}
\usepackage{bbold}
\usepackage{pgfplots}
\pgfplotsset{compat=1.16}
\pgfplotsset{every tick label/.append style={font=\small}}
\usetikzlibrary{spy,backgrounds}
\usepackage{graphicx,psfrag,epsf}
\usepackage{enumerate}
\usepackage{subcaption}
\usepackage{geometry}
\usepackage{mathtools}
\usepackage[toc]{appendix}
\usepackage{enumitem}
\usepackage[noline,ruled,linesnumbered]{algorithm2e}
\usepackage{multirow}

\title{Leave-Group-Out Cross-Validation for Latent Gaussian Models}

\author{
Zhedong Liu \\
  Statistics Program, CEMSE\\
  King Abdullah University of Science and Technology\\
  Kingdom of Saudi Arabia, Thuwal 23955-6900 \\
  \texttt{zhedongliu1@gmail.com} \\
  \And
Janet van Niekerk \\
  Statistics Program, CEMSE\\
  King Abdullah University of Science and Technology\\
  Kingdom of Saudi Arabia, Thuwal 23955-6900 \\
  Department of Statistics, University of Pretoria, South Africa\\
  \texttt{janet.vanniekerk@kaust.edu.sa} \\
   \And
 H{\aa}vard Rue \\
  Statistics Program, CEMSE\\
  King Abdullah University of Science and Technology\\
  Kingdom of Saudi Arabia, Thuwal 23955-6900 \\
  \texttt{haavard.rue@kaust.edu.sa} \\
}

\begin{document}
\maketitle
\doublespacing
\begin{abstract}
Evaluating the predictive performance of a statistical model is commonly done using cross-validation. Among the various methods, leave-one-out cross-validation (LOOCV) is frequently used. Originally designed for exchangeable observations, LOOCV has since been extended to other cases such as hierarchical models. However, it focuses primarily on short-range prediction and may not fully capture long-range prediction scenarios. For structured hierarchical models, particularly those involving multiple random effects, the concepts of short- and long-range predictions become less clear, which can complicate the interpretation of LOOCV results.
In this paper, we propose a complementary cross-validation framework specifically tailored for longer-range prediction in latent Gaussian models, including those with structured random effects. Our approach differs from LOOCV by excluding a carefully constructed set from the training set, which better emulates longer-range prediction conditions. Furthermore, we achieve computational efficiency by adjusting the full joint posterior for this modified cross-validation, thus eliminating the need for model refitting. This method is implemented in the R-INLA package (www.r-inla.org) and can be adapted to a variety of inferential frameworks.
\end{abstract}

\noindent%
{\it Keywords:}  Bayesian Cross-Validation; Latent Gaussian Models; R-INLA\\
{\it MSC:}  62-04 62C10 62F15 62J12


\section{Introduction}
\label{sec:intro}

\subsection{Rationale and Background}

Leave-one-out cross-validation (LOOCV) \cite{stone1974cross}  stands as a popular method to evaluate a statistical model's predictive performance, perform model selections, or estimating some critical parameters in the model. The core concept of LOOCV is elegantly straightforward. Suppose we have data, $\boldsymbol{y} = \{y_i\}$, for $i=1, \ldots, n$, presumed to be Independent and Identically Distributed (I.I.D.) samples from the true distribution $\pi_{T}(y)$. Our objective is to determine how well a fitted model can predict a new observation, $\tilde{y}$, sampled from this true distribution. In the Bayesian context, we use the posterior predictive distribution $\pi(y|\boldsymbol{y})$ to predict $\tilde{y}$ sampled from $\pi_{T}(y)$ as proposed by \cite{Geisser1979}. Using the logarithmic score \cite{gneiting2007strictly}, we can compute $E_{\tilde{y}}[\log \pi(\tilde{y}|\boldsymbol{y})]$ as a metric for prediction ability.

Owing to the lack of $\pi_T(y)$, directly computing the expectation becomes infeasible. Nonetheless, since
$y_i$ is an exchangeable sample from $\pi_{T}(y)$, we can estimate this expectation by evaluating $$u_{\text{LOOCV}} = \frac{1}{n} \sum_{i=1}^{n}\log {\pi}(y_i |\boldsymbol{y}_{-i}),$$ where $y_i$ is the testing point and $\boldsymbol{y}_{-i}$ is the training set, and $\boldsymbol{y}_{-i}$ are all data except the $i$th observation.

The informal interpretation of LOOCV is that it mimics ``using $\boldsymbol{y}$ to predict $\tilde{y}$" by ``using $\boldsymbol{y}_{-i}$ to predict $y_i$". This intuitive interpretation is then used to justify, often implicitly, the use of LOOCV as a ``default'' way to evaluate predictive performance.

However, issues can arise in more complex statistical models where the dependency in the model results in the data not being exchangeable (see \cite{vehtari2012survey} for a complete discussion of CV for several types of exchangeability); we describe these kinds of models as "dependent" cases for the purpose of this paper. An intuitive dependent case is a time series. \cite{burman1994cross} proposed a block CV method for dependent data from a stationary process, acknowledging the need for a different approach to CV than LOOCV. \cite{mcquarrie1998regression} propose modified cross-validation (MCV) where dependent data chunks are removed together with the relevant point to account for the dependence in a time series (and other dependent data generating models). \cite{bergmeir2012use} investigated the properties of blocked CV and other approaches for robust time series model evaluations (see also \cite{bergmeir2018note} for a study on k-fold CV), while \cite{burkner2020approximate} proposed a leave-future-out CV strategy. \cite{cerqueira2020evaluating} investigated CV and holdout approaches for time series models and concluded that the out-of-sample holdout procedure is more accurate for non-stationary processes than LOOCV.\\
Besides time series, spatial dependence models come to mind for which \cite{valavi2018blockcv} proposed a buffering strategy by leaving out specific spatial points or areas and spatial and environmental blocking. Spatial blocking forms clusters of data points according to spatial effects, and environmental blocking forms clusters using K-means \cite{hartigan1979k} on the covariates. Other examples of dependent cases are longitudinal data for multiple subjects in a study \cite{saeb2017need} and hierarchical models (see \cite{gelman1995bayesian} and \cite[Section 5.1.4.]{vehtari2012survey}). \cite{racine2000consistent} proposed an hv-block CV approach for dependent data while \cite{merkle2019bayesian} considers a multilevel model and shows that marginal WAIC is akin to LOOCV. \cite{roberts2017cross} advocate a block cross-validation, partitioning ecological data based on inherent patterns, when the prediction task is not simply short-range prediction. \cite{rabinowicz2022cross} offers a modification to LOOCV, ensuring an unbiased measure of predictive performance given the correlation between new and observed data, where the unbiasedness is in the sense of randomized both observed and new data. We should note that an assumed prediction task determines the correlation between new and observed data. \\
\\
In dependent cases, LOOCV can provide a restricted assessment of the models' predictive performance since LOOCV cannot evaluate longer-range prediction. Even in terms of short-range prediction, it is not clear what is short- or longer-range in dependent models that are not purely temporal or spatial models where the range has a physical interpretation. We use the concepts of short-range and longer-range predictions, acknowledging that these concepts can have overlapping meanings. 

We thus propose a framework that emulates longer-range prediction scenarios, for hierarchical models, by constructing non-random leave-out sets based on model-based correlations. This can be viewed as a complementary approach to LOOCV for evaluating predictive performance, providing additional informative insights of the predictive ability for dependent cases.

\subsection{The prediction task}
\label{prediction_tasks}
The critical observation is that the meaning of ``prediction" is not clearly defined when we are far away from exchangeability, so that $\boldsymbol{y}$ are \emph{non-exchangeable} samples of $\pi_T(y)$. $\pi_T(\tilde{y}|\boldsymbol{y})$ lacks a unique definition in dependent cases as without a clear \emph{prediction task}, i.e., how we imagine a new data point, $\tilde{y}$, is generated given observed data $\boldsymbol{y}$. This ambiguity extends to the act of ``using $\boldsymbol{y}$ to predict $\tilde{y}$" as it is uncertain what our target, $\tilde{y}$, represents. To illustrate these concepts, let us discuss some more concrete examples.

\subsubsection*{Time-series model} Assume data $\boldsymbol{y} = \{y_1, y_2, \dots, y_T\}$ is a time-series, observed sequentially at time $1, 2, \dots, T$. The inherent prediction task is to predict future values, given the temporal nature of the data. We can predict a new observation at $k\ge 1$ steps into the future by $\pi(y_{T + k} | y_1, \dots, y_T)$.

In this example, the LOOCV will be computed from 
\begin{displaymath}
    \pi(y_t | y_1, \ldots, y_{t-1}, y_{t+1}, \dots, y_T), \qquad t=1, \ldots, T,
\end{displaymath}
which is often referred to as interpolation or imputation of missing values, rather than a prediction. However, the predictive performance of time series models is often assessed through leave-future-out cross-validation (LFOCV) \cite{burkner2020approximate}:
$$
    \sum_{T'=T_0}^{T-k} \log\pi(y_{T' + k} | y_1, \dots, y_{T'}),
$$
where $T'$ starts from time $T_0>1$ as we need some data to estimate the model.

The message from this example is that LOOCV, when applied to such models, is essentially evaluating short-range prediction performance rather than longer-range predictive performance.

We acknowledge two issues. First, the distinction between short and longer-range prediction is not always clear-cut, leading to overlapping concepts. For example, a one-step-ahead forecast leans more towards short range than a two-step-ahead prediction. In contrast, a one-step-ahead forecast leans less towards short-range than a missing value imputation. However, this does not deter our discussion. Secondly, while an ideal model succeeds in all prediction tasks, real-world scenarios require us to settle for the definition of the ``best fit". Consequently, our choice of evaluation should align with our specific objectives.

\subsubsection*{Multilevel model}
\begin{figure}
    \resizebox{\textwidth}{!}{
        \begin{forest} [$\mathcal{D}$ [$r_1$ [$s_1$ [$c_1$ [$y_1$]
            [$y_2$] [$y_3$]] [$c_2$ [$y_4$] [$y_5$]]] [$s_2$ [$c_3$
            [$y_6$] [$y_7$]] [$c_4$ [$y_8$] [$y_9$]]]] [$r_2$ [$s_3$
            [$c_5$ [$y_{10}$] [$y_{11}$]] [$c_6$ [$y_{12}$] [$y_{13}$]
            [$y_{14}$] [$y_{15}$]]] [$s_4$ [$c_7$ [$y_{16}$]] [$c_8$
            [$y_{17}$] [$y_{18}$]]]]]
        \end{forest}
    }
    \caption{A nested multilevel model.}
    \label{fig:data_school}
\end{figure}
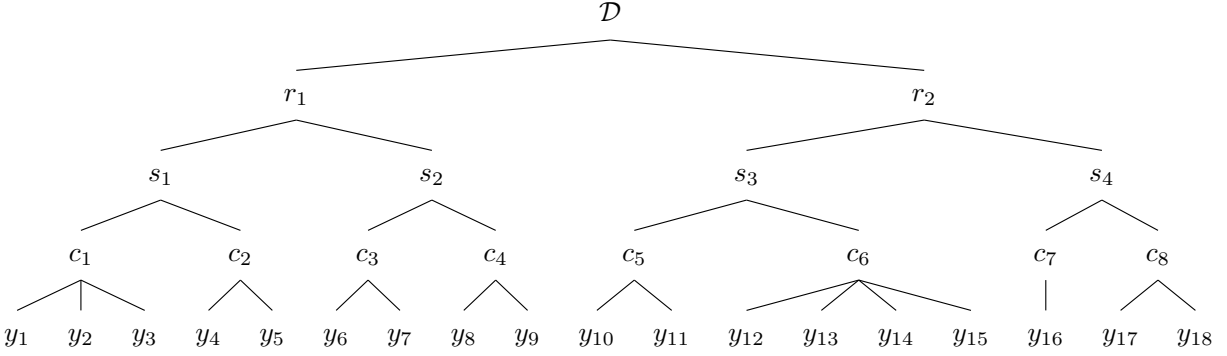

Figure \ref{fig:data_school} illustrates an example of a multilevel model. Consider observations of student grades or performance. This data exhibits a hierarchical structure: students belong to classes, classes reside within schools, and schools are nested within regions. This hierarchical arrangement is significant because it introduces correlated random effects attributed to the class, school, and region levels, substantially deviating from the exchangeable case.

Given such a model, the prediction task becomes ambiguous. Are we aiming to predict the performance of an unobserved student from an observed class? Or are we trying to predict the performance of an unobserved student in an unobserved class, school, or even region? This difficulty mirrors the challenges in defining asymptotic regimes for these models. As students, classes, schools, and regions can grow indefinitely in various ways, it is unclear whether one of such choices is the most reasonable.

To evaluate predictive performance within this context, users must first explicitly define their prediction task and then evaluate the model according to this definition. It should be noted that applying LOOCV would evaluate the prediction of individual students within observed classes. In our view, this mimics more short-range prediction rather than longer-range prediction, and another framework is needed to quantify the predictive ability for a new student in a new class in a new school in a new region, for example. Our proposal provides some insight into this kind of prediction task.

\subsection{LGOCV: Complementing LOOCV for dependent cases}

Our discussions illuminate an important insight: when dealing with models that lead to non-exchangeable data, the prediction task implicitly defined through LOOCV may be less appropriate, as it leans more towards assessing imputing qualities and short range predictions than predictive performance for longer range as is usually implied by "out-of-sample" prediction. This prompts the question: What is a suitable approach moving forward?

One observation is the absence of a ``one size fits all" solution. Each model may possess a natural prediction task—or several—based on its intended application. Thus, for a specific assessment of predictive performance, we need to define these prediction tasks explicitly. One can then evaluate distinct predictive performance metrics using our proposed leave-group-out cross-validation (LGOCV):
\begin{equation}
\label{u_LGOCV}
    u_{\text{LGOCV}} = \frac{1}{n}\sum_{i=1}^n \log (\pi(y_i|\boldsymbol{y}_{-I_i})).
\end{equation}
Here, the \emph{group} (denoted by $I_i$) is an index set including $i$. This configuration facilitates that the pair $(y_i,\boldsymbol{y}_{-I_i})$ mimics a specified prediction task, with $\boldsymbol{y}_{-I_i}$ being the data subset excluding the data indexed by $I_i$. In a multilevel model, as depicted in Figure \ref{fig:data_school}, predicting a student's grade from an unseen class necessitates that $I_i$ includes $i$ and all observations from student $i$'s class. However, more complex models, such as models containing both time series and hierarchical elements, pose challenges when defining a natural prediction task.  Therefore, even in complex cases, LOOCV is often applied for its simplicity—even if it leans more towards imputation or short-range prediction.

Developing a framework that evaluates a model's longer-range prediction like the proposed LGOCV, necessitates the construction of the leave-out group $I_i$ for each datapoint $y_i$. Our approach constructs a model-based group, $I_i$, for each $i$ by using the prior or posterior correlation among the set of linear predictors. Though we will delve into the construction of $I_i$ in Section \ref{sec:auto}, an initial understanding is that $I_i$ comprises the data points that correspond to the linear predictors that are most informative for predicting the testing linear predictor, and thus the testing point, $y_i$. This set ensures that our LGOCV focuses less on short-range prediction (interpolation) and more on longer-range prediction than LOOCV. In other words, LGOCV tests the model on more difficult prediction tasks since the most influential points are removed together with the testing point, instead of some arbitrary (possibly uncorrelated) point(s). The user needs to only provide a number that indicates the "degree of the independence" between the prediction point and the rest of the data", and we compute these groups for each datapoint in an automated way. In various practical examples, we will show how this model-based procedure produces reasonable groups. Advanced spatial examples applying the proposed method are presented by \cite{adin2024automatic}. For a simple time-series example, our new approach will correspond to evaluating $\pi(y_t | y_1, \ldots, y_{t-k}, y_{t+k}, \ldots, y_T), $
for fixed $k>1$. This corresponds to removing a sequence of data with length $2k-1$, to predict the central one. As we see, this task mimics a longer-range prediction task. Our interpretation is that LGOCV quantifies the model's ability to predict longer-range more appropriately than LOOCV, when $k > 1$, and is similar to the cross-validation procedure proposed by \cite{burman1994cross} for stationary processes.

There are two key challenges to address to make our proposal practical.  Firstly, we must quantify the information contributed by one data point in predicting another; this is crucial for the group construction. Secondly, we face the computational task of evaluating $u_{\text{LGOCV}}$ given a set of groups. The naive computation of LGOCV by fitting models across all potential training sets and evaluating their utility against corresponding testing points is computationally infeasible, especially given the resource-demanding nature of modern statistical models. However, these challenges can be handled elegantly within the framework of latent Gaussian models (LGMs) combined with the integrated nested Laplace approximation (INLA) inference, as detailed in \cite{rue2009approximate,rue2017bayesian,van2021correcting,van2023new}.  Throughout this paper, we will assume that our model is an LGM. We will discuss how to integrate the automatic group construction and the fast computation of $u_{\text{LGOCV}}$ using the INLA framework. Notably, our proposed methodology has been incorporated into the R-INLA package (www.rinla.org), extending its applicability across all LGMs supported by R-INLA.

\subsection{Theoretical aspects}

Cross-validation (CV), particularly LOOCV, is frequently considered as an estimator of $E_{\tilde{y}}[\log \pi(\tilde{y}|\boldsymbol{y})]$ or $E_{\tilde{y},\boldsymbol{y}}[\log \pi(\tilde{y}|\boldsymbol{y})]$. The first expectation describes the generalized predictive performance given a specific training set, while the second expectation describes the generalized predictive performance averaged over different identically distributed training sets. These expectations can be evaluated when assuming the existence of the joint density $\pi_{T}(\tilde{y}, \boldsymbol{y})$, representing the true data generation process. Under the assumption of exchangeability and some regularity conditions on the model, the Bernstein-Von-Mises theorem states that  $\log \pi(\tilde{y}|\boldsymbol{y})$ converges to a random variable irrelevant to $\boldsymbol{y}$. Consequently, $E_{\tilde{y}}[\log \pi(\tilde{y}|\boldsymbol{y})]$ and $E_{\tilde{y},\boldsymbol{y}}[\log \pi(\tilde{y}|\boldsymbol{y})]$ become equivalent in the limit. If we further assume that $\tilde{y}$ is sampled from the same distribution as all the training data, LOOCV is an asymptotically unbiased estimator of the expectations. Commonly used information criteria, such as AIC \cite{akaike1973information}, WAIC \cite{watanabe2010asymptotic}, are asymptotically equivalent to LOOCV in fully exchangeable cases. This type of analysis is prevalent in the literature with various settings \cite{stone1974cross,stone1977asymptotic,yang2007consistency,shao1993linear}.

However, a similar analysis does not hold for dependent cases in general. Firstly, the existence of different prediction tasks means that both the model prediction, $\pi(\tilde{y}|\boldsymbol{y})$, and the true data generation process, $\pi_{T}(\tilde{y}|\boldsymbol{y})$, are not uniquely defined as discussed in Section \ref{prediction_tasks}. Secondly, the asymptotic scheme is not uniquely defined, even with a specific prediction task. For example, in a temporal model where data $\boldsymbol{y} = \{y_1, y_2, \dots, y_n\}$ is a time-series, observed at time $t_1< t_2< \dots < t_n$ and we denote the last time step as $T$. Several meanings of of $n \to \infty$ can be considered:
\begin{itemize}
    \item $T\to \infty$ and $t_i - t_{i-1}$ is a constant
    \item $t_i - t_{i-1} \to 0$ and $T$ is a constant
    \item $t_i - t_{i-1} \to 0$ and $T \to \infty$ with $T(t_i - t_{i-1})$ fixed
\end{itemize}
These scenarios correspond to observing more future data and having higher sample rates within a time frame. As mentioned in Section \ref{prediction_tasks}, multilevel data can also have various asymptotic regimes. Thirdly, if the data generation process is not stationary, the model will not converge under certain asymptotic regimes, which differentiate $E_{\tilde{y}}[\log \pi(\tilde{y}|\boldsymbol{y})]$ from $E_{\tilde{y},\boldsymbol{y}}[\log \pi(\tilde{y}|\boldsymbol{y})]$ even in asymptotic scenarios. These points highlight that the estimand of CV is not uniquely defined in dependent cases, preventing the establishment of an asymptotic analysis framework.

From the perspective of CV, it is also inappropriate to consider it an estimator since each summand in CV should be viewed as a sample from different distributions due to the relevance of data indexes in dependent cases. For example, if we compute LOOCV in a time series. Each $y_t$ is sampled from a different conditional distribution $\pi_{T}(y_t|\boldsymbol{y}_{-t})$ and thus the average $\frac{1}{n} \sum_{t=1}^T \log \pi(y_t|\boldsymbol{y}_{-t})$ cannot be considered as an estimator in general. Therefore, it is more reasonable to view CV as a predictive measurement rather than an estimator of an expectation. This perspective allows us to interpret the proposed LGOCV as the averaged predictive performance for similar prediction tasks, created systematically by the model.

While the proposal of \cite{merkle2019bayesian} for multilevel model demonstrates that marginal WAIC is akin to LOOCV, we note that conditional WAIC aligns with LGOCV, where a hierarchical level, such as a school, defines the groups. The h-block CV of \cite{burman1994cross} is a special case of LGOCV for a stationery model. LFOCV proposed by \cite{burkner2020approximate} is similar to LGOCV as shown in Section \ref{sec:verify}. The spatial buffering proposed by \cite{valavi2018blockcv} ensures that no test data is spatially next to any training data, and is a special case of LGOCV for model with only spatial effects. LGOCV this provides a framework where no training data is placed next to the test data in terms of the entire model, and not just specific components thereof. 

\subsection{Plan of paper}

We propose the model-based LGOCV to evaluate longer-range prediction performance for latent Gaussian models, as a special case of a hierarchical model. Complementing this, we introduce a computational method to approximate $u_{\text{LGOCV}}$ without model refitting, which is crucial for practical implementation of our proposal. Our computational technique also facilitates the calculation of $u_{\text{LGOCV}}$ with user-specified groups.

Section \ref{sec:LGMs} introduces LGMs and explains how they can be efficiently inferred using INLA. In Section \ref{sec:auto}, we discuss the model-based group construction method for LGMs. This method can be implemented in two ways: by using the prior correlation matrix or the posterior correlation matrix of the latent linear predictors. In Section \ref{sec:method}, we demonstrate how to approximate the LGOCV predictive density. Finally, in Section \ref{sec:verify}, we compare the approximated LGOCV with the exact LGOCV computed by Markov chain Monte Carlo (MCMC) and present some applications. We conclude with a general discussion in Section \ref{sec:conc}.
 
\section{Latent Gaussian models}
\label{sec:LGMs}
This section briefly introduces LGMs, as detailed in \cite{rue2009approximate,rue2017bayesian,van2021correcting,van2023new}, since the model-based group construction and fast approximations rely on them. The LGMs can be formulated by
\begin{equation}
\label{eq:lgm}
    \begin{split}
        &y_i|\eta_i,\boldsymbol{\theta} \sim \pi(y_i|\eta_i,\boldsymbol{\theta}), \\
        &\boldsymbol{\eta} = \boldsymbol{A}\boldsymbol{f}, \quad\boldsymbol{f}|\boldsymbol{\theta} \sim N(0,\boldsymbol{P}_{\boldsymbol{f}}(\boldsymbol{\theta})), \quad\boldsymbol{\theta} \sim \pi(\boldsymbol{\theta}).
    \end{split}
\end{equation}
In LGMs, each $y_i$ is independent conditioned on its corresponding linear predictor $\eta_i$, and hyperparameters $\boldsymbol{\theta}$; $\boldsymbol{\eta}$ is a linear combination of $\boldsymbol{f}$, which is assigned with a Gaussian prior with zero mean and a precision matrix parameterized by $\boldsymbol{\theta}$; $\boldsymbol{A}$ is the design matrix mapping $\boldsymbol{f}$ to $\boldsymbol{\eta}$; $\pi(\boldsymbol{\theta})$ is a prior density of hyperparameters. It is worth mentioning that the prior precision matrix $\boldsymbol{P}_{\boldsymbol{f}}(\boldsymbol{\theta})$ is very sparse, which is leveraged to speed up the inference. 

The model is quite general because $\boldsymbol{f}$ can combine many modeling components, including linear model, spatial components, temporal components, spline components, etc \cite{wang2018bayesian,krainski2018advanced,gomez2020bayesian}. It is also common with linear constraints on the latent effects $\boldsymbol{f}$ \cite{rue2005gaussian}.

We can approximate $\pi(\boldsymbol{f}|\boldsymbol{\theta},\boldsymbol{y})$ and $\pi(\boldsymbol{\theta}|\boldsymbol{y})$ at some configurations, $\boldsymbol{\theta}_1 \dots \boldsymbol{\theta}_k$. The configurations are located around the mode of $\pi(\boldsymbol{\theta}|\boldsymbol{y})$, denoted by $\boldsymbol{\theta}^*$, for numerical integration. Approximations of $\pi(\boldsymbol{\eta}|\boldsymbol{\theta},\boldsymbol{y})$ are computed using the linear relation, $\boldsymbol{\eta} = \boldsymbol{A}\boldsymbol{f}$. The Gaussian approximation of $\pi(\boldsymbol{f}|\boldsymbol{\theta},\boldsymbol{y})$ plays an essential role, which is outlined as follows.

We have $\pi(\boldsymbol{f}|\boldsymbol{\theta},\boldsymbol{y})$ for a given $\boldsymbol{\theta}$,
\begin{equation}
    \label{eq:prop_beta}
    \pi(\boldsymbol{f}|\boldsymbol{\theta},\boldsymbol{y}) \propto \exp\bigg\{-\frac{1}{2}\boldsymbol{f}^T\boldsymbol{P}_{\boldsymbol{f}}(\boldsymbol{\theta})\boldsymbol{f} + \sum_{i=1}^n \log(\pi(y_i|\eta_i,\boldsymbol{\theta}))\bigg\},
\end{equation}
whose mode is $\boldsymbol{\mu}_{\boldsymbol{f}}(\boldsymbol{\theta},\boldsymbol{y})$. The Gaussian approximation of $\pi(\boldsymbol{f}|\boldsymbol{\theta},\boldsymbol{y})$ is 
\begin{equation}
\label{eq:beta_gaussian_matrix}
\pi_G(\boldsymbol{f}|\boldsymbol{\theta},\boldsymbol{y})\propto \exp\bigg\{-\frac{1}{2}\boldsymbol{f}^T(\boldsymbol{P}_{\boldsymbol{f}}(\boldsymbol{\theta}) + \boldsymbol{A}^T\boldsymbol{C}(\boldsymbol{\theta},\boldsymbol{y})\boldsymbol{A})\boldsymbol{f} + \boldsymbol{A}^T\boldsymbol{b}(\boldsymbol{\theta},\boldsymbol{y})\boldsymbol{f}\bigg\}.
\end{equation}
In (\ref{eq:beta_gaussian_matrix}), $b_i(\boldsymbol{\theta},\boldsymbol{y}) = g_i'(\eta_i^*) - g_i''(\eta_i^*)\eta_i^*$, and $\boldsymbol{C}(\boldsymbol{\theta},\boldsymbol{y})$ is a diagonal matrix with $C_{ii}(\boldsymbol{\theta},\boldsymbol{y}) = - g_i''(\eta_i^*)$, where $g_i(\eta_i) = \log(\pi(y_i|\eta_i,\boldsymbol{\theta}))$ and $\eta_i^* = \boldsymbol{A}_i\boldsymbol{\mu}_{f}(\boldsymbol{\theta},\boldsymbol{y})$ with $\boldsymbol{A}_i$ being $i$th row of $\boldsymbol{A}$. The Gaussian approximation is denoted by,
\begin{equation}
    \label{eq:beta_gaussian}
    \boldsymbol{f}|\boldsymbol{y},\boldsymbol{\theta} \approx N(\boldsymbol{\mu}_{\boldsymbol{f}}(\boldsymbol{\theta},\boldsymbol{y}),\boldsymbol{Q}_{\boldsymbol{f}}(\boldsymbol{\theta},\boldsymbol{y})),
\end{equation}
where $\boldsymbol{\mu}_{f}(\boldsymbol{\theta},\boldsymbol{y}) = \boldsymbol{Q}_{\boldsymbol{f}}(\boldsymbol{\theta},\boldsymbol{y})^{-1}\boldsymbol{A}^T\boldsymbol{b}(\boldsymbol{\theta},\boldsymbol{y})$ and $\boldsymbol{Q}_{\boldsymbol{f}}(\boldsymbol{\theta},\boldsymbol{y}) = \boldsymbol{P}_{\boldsymbol{f}}(\boldsymbol{\theta}) + \boldsymbol{A}^T\boldsymbol{C}(\boldsymbol{\theta},\boldsymbol{y})\boldsymbol{A}$ are the approximated posterior mean and precision matrix of $\boldsymbol{f}$ given $\boldsymbol{\theta}$.

\section{Model-based Group Construction}
\label{sec:auto}
The primary feature of our proposed group construction is that it requires a choice of correlation matrix (prior or posterior) for the linear predictors, and a single mandatory parameter to adjust the difficulty of the prediction task. This parameter is termed "the number of level sets". It can be interpreted as the strength of the non-dependence between the group to leave out and the rest of the data. A higher value would thus ensure that the leave-out group is more independent from the rest of the data, than a lower value. A higher indpeendnece between the leave-out data and the rest of the data simluates a more difficult prediction task for the model. Based on this value and the correlation matrix choice, all other processes are automated. In a multivariate Gaussian distribution, we can quantify the information provided by a data point to predict another data point by the variance reduction of the conditional distribution, and the variance reduction is a function of their correlation coefficient. To elaborate, if $X$ and $Y$ are both Gaussian random variables, the variance reduction resulting from knowing $X$ when predicting $Y$ equates to $\sigma^2_{Y} \rho^2$, where $\sigma^2_{Y}$ is the marginal variance of $Y$ and $\rho$ is the correlation between $X$ and $Y$.
    
In LGMs, the linear predictors, $\boldsymbol{\eta}$, represent the underlining data generation process of data in (\ref{eq:lgm}). The linear predictors are designed to have a Gaussian prior and approximated to be Gaussian in posterior given $\boldsymbol{\theta}$ therefore, we can use the absolute value of the correlation matrix of $\boldsymbol{\eta}$ to represent the information provided by one data point to predict another data point. We evaluate those correlation matrices at the mode of $\pi(\boldsymbol{\theta}|\boldsymbol{y})$, denoted by $\boldsymbol{\theta}^*$. Then, we have correlation matrices of  $\boldsymbol{\eta}$ derived from the prior precision matrix, $\boldsymbol{P}_{\boldsymbol{f}}(\boldsymbol{\theta}^*)$, and the posterior precision matrix, $\boldsymbol{Q}_{\boldsymbol{f}}(\boldsymbol{\theta}^*,\boldsymbol{y})$. We call the former one prior correlation matrix, denoted by $\boldsymbol{R}_{\text{prior}}$, and the latter one posterior correlation matrix, denoted by $\boldsymbol{R}_{\text{post}}$. Note that the correlation matrices are not fully evaluated and stored to avoid computational burden as they are dense and large; thus, care has to be applied to the implementation to make it feasible. The group would vary with $\boldsymbol{\theta}$. We use $\boldsymbol{\theta}^*$ because it has the highest weight in the posterior. This preference arises because we frequently employ non-informative priors for hyperparameters. This approach ensures that our focus remains on evidence from the data rather than on arbitrary assumptions about hyperparameters.

Manually constructed groups are often based on prior knowledge and some structured effects, represented by $\boldsymbol{f}$. To imitate this process, we can compute the correlation matrix from a submatrix of $\boldsymbol{P}_{\boldsymbol{f}}(\boldsymbol{\theta})$. The correlation matrix, $\boldsymbol{R}_{\text{prior}}$, derived from the submatrix of the prior precision matrix, is a correlation matrix conditioning on those unselected effects. The groups constructed using $\boldsymbol{R}_{\text{prior}}$ are viewed to be solely user-defined in the way that it only depends on the priors and not on the data. In some situations this could be motivated, but in general we recommend using $\boldsymbol{R}_{\text{post}}$ to construct groups because the data will be informative to determine the importance of each effect. 

When using a correlation matrix $\boldsymbol{R}$, it is natural to select a fixed number of $\eta_j$ most correlated to $\eta_i$ and include their index in the group $I_i$. However, this approach can be problematic as some linear predictors may have identical absolute correlations to $\eta_i$, e.g., in a model with only intercept, all the linear predictors are correlated to each other with correlation $1$. Instead, we include all indices of $\eta_j$'s with identical absolute correlations to $\eta_i$ in $I_i$ if one of them is included. We define a level set as all $\eta_j$'s with the same absolute correlation to $\eta_i$ and determine the group size based on the number of level sets, denoted as $m$. Setting a higher $m$ results in a less dependent training set and testing point. We recommend using a small number of level sets, such as $m=3$, as a high value of $m$ can result in a large leave-out group size.

The automated group construction process thus involves selecting the number of level sets, $m$, and the correlation matrix to use, $\boldsymbol{R}_{\text{prior}}$ or $\boldsymbol{R}_{\text{post}}$. For each $i$, we can associate $m$ level sets with the $m$ largest absolute correlations to $\eta_i$, and the union of those level sets forms $I_i$. As an illustration, we outline the automated group construction procedure in Algorithm \ref{algo1}.


\begin{algorithm}[H]
\label{algo1}
\SetAlgoLined
\textbf{Input:} A correlation matrix choice $\boldsymbol{R}$ (posterior correlation is the default), Number of level sets $m$\;
\textbf{Output:} A list containing the groups for all data points\;
Calculate $\boldsymbol{R}$ from the model\;
$N \leftarrow \text{number of rows in } \boldsymbol{R}$\;
$\text{groups} \leftarrow \text{initialize } N \text{ empty lists}$\;
\For{$i=1$ \KwTo $N$}{
    $\boldsymbol{r} \leftarrow \text{absolute values of the } i\text{-th row of } \boldsymbol{R}$\;
    $\text{ordered indices} \leftarrow \text{indices of } \boldsymbol{r}  \text{ sorted by value in decreasing order}$\;
    $\text{current absolute correlation} \leftarrow 1$\;
    $k \leftarrow 1$ \;
    \For{$j=1$ \KwTo $m$}{
        \While{$\text{current absolute correlation} == \boldsymbol{r}[\text{ordered indices}[k]]$}{
            \text{groups}[i].append(\text{ordered indices}[k])\;
            $k \leftarrow k + 1$\;
        }
        $\text{current absolute correlation} \leftarrow \boldsymbol{r}[\text{ordered indices}[k]]$\;
    }
}
\Return{$\text{groups}$}\;
\caption{Find groups for all data points}
\end{algorithm}
     
\section{Approximation of LGOCV predictive density}
\label{sec:method}
In this section, we will explore the process of approximating $\pi(y_i|\boldsymbol{y}_{-I_i})$. The results are straightforward but tedious in implementation; thus, it is crucial to exercise caution to ensure that all potential numerical instabilities are accounted for. Through empirical testing, this new method has shown to be both more accurate and stable compared to the approach outlined in \cite{rue2009approximate}, when $I_i = i$.

We start by writing $\pi(y_i|\boldsymbol{y}_{-I_i})$ as nested integrals,
\begin{align}
	\label{eq:nest_int_hyper}
	& \pi(y_i|\boldsymbol{y}_{-I_i}) = \int_{\boldsymbol{\theta}}\pi(y_i|\boldsymbol{\theta},\boldsymbol{y}_{-I_i})\pi(\boldsymbol{\theta}|\boldsymbol{y}_{-I_i})d\boldsymbol{\theta}\\
	\label{eq:nest_int_linear}
    & \pi(y_i|\boldsymbol{\theta},\boldsymbol{y}_{-I_i}) = \int \pi(y_i|\eta_i,\boldsymbol{\theta})\pi(\eta_i|\boldsymbol{\theta},\boldsymbol{y}_{-I_i})d\eta_i.
\end{align}
The integral (\ref{eq:nest_int_hyper}) is computed by the numerical integration \cite{rue2009approximate}, and the integral (\ref{eq:nest_int_linear}) is computed by Gauss-Hermite quadratures \cite{liu1994note} as the \emph{conditional} posterior density $\pi(\eta_i|\boldsymbol{\theta},\boldsymbol{y}_{-I_i})$ will be approximated by a Gaussian distribution. The key to the accuracy of (\ref{eq:nest_int_linear}) is that the likelihood, $\pi(y_i|\eta_i,\boldsymbol{\theta})$, is known such that small approximation errors of $\pi(\eta_i|\boldsymbol{\theta},\boldsymbol{y}_{-I_i})$ diminish due to the integration. The accuracy of (\ref{eq:nest_int_hyper}) relies on the accuracy of (\ref{eq:nest_int_linear}) and the assumption that the removal of $\boldsymbol{y}_{I_i}$ does not have a dramatic impact on the posterior.

The computation of the nested integrals reduces to the computation of $\pi(\eta_i|\boldsymbol{\theta},\boldsymbol{y}_{-I_i})$ and $\pi(\boldsymbol{\theta}|\boldsymbol{y}_{-I_i})$. We will approximate $\pi(\eta_i|\boldsymbol{\theta},\boldsymbol{y}_{-I_i})$ by a Gaussian distribution, denoted by $\pi_G(\eta_{i}|\boldsymbol{\theta},\boldsymbol{y}_{-I_i})$, and $\pi(\boldsymbol{\theta}|\boldsymbol{y}_{-I_i})$ by correcting the approximation of $\pi(\boldsymbol{\theta}|\boldsymbol{y})$ in \cite{rue2009approximate}. We further improve the mean of $\pi_G(\eta_i|\boldsymbol{\theta},\boldsymbol{y}_{-I_i})$ using variational Bayes \cite{van2021correcting} in the implementation. In this section, we focus on the explanation of computing $\pi_G(\eta_{i}|\boldsymbol{\theta},\boldsymbol{y}_{-I_i})$ and an approximation of $\pi(\boldsymbol{\theta}|\boldsymbol{y}_{-I_i})$.

\subsubsection*{Computing $\pi_G(\eta_{i}|\boldsymbol{\theta},\boldsymbol{y}_{-I_i})$}
The mean and variance of $\pi_G(\eta_{i}|\boldsymbol{\theta},\boldsymbol{y}_{-I_i})$ can be obtained by 
\begin{equation}
\label{naive}
\
	\begin{split}
	&\mu_{\eta_i}(\boldsymbol{\theta},\boldsymbol{y}_{-I_i}) = \boldsymbol{A}_i\boldsymbol{\mu}_{
\boldsymbol{f}}(\boldsymbol{\theta},\boldsymbol{y}_{-I_i}), \\
	&\sigma^2_{\eta_i}(\boldsymbol{\theta},\boldsymbol{y}_{-I_i}) = \boldsymbol{A}_i\boldsymbol{Q}_{
\boldsymbol{f}}^{-1}(\boldsymbol{\theta},\boldsymbol{y}_{-I_i})\boldsymbol{A}_i^T.
	\end{split}
\end{equation}
The computation of $\pi_G(\boldsymbol{f}|\boldsymbol{\theta},\boldsymbol{y}_{-I_i})$ requires the mode of $\pi(\boldsymbol{f}|\boldsymbol{\theta},\boldsymbol{y}_{-I_i})$ for each $i$ at each configuration of $\boldsymbol{\theta}$, which is computationally expensive. With the mode at full data, we use an approximation to avoid the optimization step,
\begin{align}
\label{approx_matrix}
	&\boldsymbol{Q}_{\boldsymbol{f}}(\boldsymbol{\theta},\boldsymbol{y}_{-I_i}) \approx \tilde{\boldsymbol{Q}}_{\boldsymbol{f}}(\boldsymbol{\theta},\boldsymbol{y}_{-I_i})= \boldsymbol{Q}_{\boldsymbol{f}}(\boldsymbol{\theta},\boldsymbol{y}) - \boldsymbol{A}_{I_i}^T\boldsymbol{C}_{I_i}(\boldsymbol{\theta},\boldsymbol{y})\boldsymbol{A}_{I_i}  ,\\
\label{approx_mean}
	&\boldsymbol{\mu}_{\boldsymbol{f}}(\boldsymbol{\theta},\boldsymbol{y}_{-I_i}) \approx \tilde{\boldsymbol{\mu}}_{\boldsymbol{f}}(\boldsymbol{\theta},\boldsymbol{y}_{-I_i})= \tilde{\boldsymbol{Q}}_{\boldsymbol{f}}(\boldsymbol{\theta},\boldsymbol{y}_{-I_i})^{-1}(\boldsymbol{A}^T\boldsymbol{b}(\boldsymbol{\theta},\boldsymbol{y}) - \boldsymbol{A}_{I_i}^T\boldsymbol{b}_{I_i}(\boldsymbol{\theta},\boldsymbol{y})),
\end{align}
where $\boldsymbol{A}_{I_i}$ is a submatrix of $\boldsymbol{A}$ formed by rows of $\boldsymbol{A}$, $\boldsymbol{b}_{I_i}(\boldsymbol{\theta},\boldsymbol{y})$ is a subvector of $\boldsymbol{b}(\boldsymbol{\theta},\boldsymbol{y})$, and $\boldsymbol{C}_{I_i}(\boldsymbol{\theta},\boldsymbol{y})$ is a principal submatrix of $\boldsymbol{C}(\boldsymbol{\theta},\boldsymbol{y})$. When the posterior is Gaussian, the approximation is exact as (\ref{approx_matrix}) and (\ref{approx_mean}) define the precision matrix and the mean of the posterior. It seems easy to obtain the moments using (\ref{naive}), but the decomposition of $\tilde{\boldsymbol{Q}}_{\boldsymbol{f}}(\boldsymbol{\theta},\boldsymbol{y}_{-I_i})$ is too expensive. To avoid the decomposition of $\tilde{\boldsymbol{Q}}_{\boldsymbol{f}}(\boldsymbol{\theta},\boldsymbol{y}_{-I_i})$, we use the linear relation
$\boldsymbol{\eta}_{I_i} = \boldsymbol{A}_{I_i}\boldsymbol{f}$ to map all the computation on $\boldsymbol{f}$ to $\boldsymbol{\eta}_{I_i}$. We compute $\boldsymbol{\Sigma}_{\boldsymbol{\eta}_{I_i}}(\boldsymbol{\theta},\boldsymbol{y}_{-I_i})$ and $\boldsymbol{\mu}_{\boldsymbol{\eta}_{I_i}}(\boldsymbol{\theta},\boldsymbol{y}_{-I_i})$ through $\boldsymbol{\Sigma}_{\boldsymbol{\eta}_{I_i}}(\boldsymbol{\theta},\boldsymbol{y})$ and $\boldsymbol{\mu}_{\boldsymbol{\eta}_{I_i}}(\boldsymbol{\theta},\boldsymbol{y})$ as shown in the Appendix A using a low rank representation, where $\boldsymbol{\Sigma}_{\boldsymbol{\eta}_{I_i}}(\boldsymbol{\theta},\boldsymbol{y})$ is the posterior covariance matrix of $\boldsymbol{\eta}_{I_i}$ and $\boldsymbol{\Sigma}_{\boldsymbol{\eta}_{I_i}}(\boldsymbol{\theta},\boldsymbol{y}_{-I_i})$ is the covariance matrix of $\boldsymbol{\eta}_{I_i}$ with $\boldsymbol{y}_{I_i}$ left out. The computation of $\boldsymbol{\Sigma}_{\boldsymbol{\eta}_{I_i}}(\boldsymbol{\theta},\boldsymbol{y})$ is non-trivial, especially when linear constraints are applied, which is demonstrated in Appendix B.

The approximation is more accurate when $\pi(\eta_{i}|\boldsymbol{\theta},\boldsymbol{y}_{-I_i})$ is close to Gaussian. The Gaussianity of $\eta_{i}|\boldsymbol{\theta},\boldsymbol{y}_{-I_i}$ comes from three sources. Firstly, $\pi(\eta_{i}|\boldsymbol{\theta},\boldsymbol{y}_{-I_i})$ is nearly Gaussian, when $\eta_{i}$ is connected to large amount of data \cite{rue2009approximate}. Secondly, $\pi(\eta_{i}|\boldsymbol{\theta},\boldsymbol{y}_{-I_i})$ is dominated by the Gaussian prior, which happens when $\eta_{i}$ is connected to very few data. Thirdly, the log-likelihood can be close to the log-likelihood of a Gaussian distribution, resulting in the Gaussianity of $\pi(\eta_{i}|\boldsymbol{\theta},\boldsymbol{y}_{-I_i})$ due to the conjugacy. Thus, $\pi(\eta_{i}|\boldsymbol{\theta},\boldsymbol{y}_{-I_i})$ is rarely far away from a Gaussian distribution.

\subsubsection*{Approximating $\pi(\boldsymbol{\theta}|\boldsymbol{y}_{-I_i})$}
To approximate $\pi(\boldsymbol{\theta}|\boldsymbol{y}_{-I_i})$, we use the relation,$\pi(\boldsymbol{\theta}|\boldsymbol{y}_{-I_i}) \propto \frac{\pi(\boldsymbol{\theta}|\boldsymbol{y})}{\pi(\boldsymbol{y}_{I_i}|\boldsymbol{\theta},\boldsymbol{y}_{-I_i})},$ where we can approximate $\pi(\boldsymbol{\theta}|\boldsymbol{y})$ at configurations as in \cite{rue2009approximate}. We need to compute
$\pi(\boldsymbol{y}_{I_i}|\boldsymbol{\theta},\boldsymbol{y}_{-I_i}) \approx \int \pi(\boldsymbol{y}_{I_i}|\boldsymbol{\eta}_{I_i},\boldsymbol{\theta})\pi_G(\boldsymbol{\eta}_{I_i}|\boldsymbol{\theta},\boldsymbol{y}_{-I_i})d\boldsymbol{\eta}_{I_i}.$
A Laplace approximation can be applied to this integral,
\begin{equation}
\label{LA_theta_correction}
\pi_{\text{LA}}(\boldsymbol{y}_{I_i}|\boldsymbol{\theta},\boldsymbol{y}_{-I_i}) = \frac{\pi(\boldsymbol{y}_{I_i}|\boldsymbol{\eta}^*_{I_i},\boldsymbol{\theta})\pi_G(\boldsymbol{\eta}^*_{I_i}|\boldsymbol{\theta},\boldsymbol{y}_{-I_i})}{\pi_G(\boldsymbol{\eta}^*_{I_i}|\boldsymbol{\theta},\boldsymbol{y})},
\end{equation}
where $\boldsymbol{\eta}^*_{I_i}$ is the mode of $\pi_G(\boldsymbol{\eta}^*_{I_i}|\boldsymbol{\theta},\boldsymbol{y})$. Note that the correction of the hyperparameter reuses $\pi_G(\boldsymbol{\eta}_{I_i}|\boldsymbol{\theta},\boldsymbol{y}_{-I_i})$ and $\pi_G(\boldsymbol{\eta}_{I_i}|\boldsymbol{\theta},\boldsymbol{y})$.

\section{Simulations and Applications}
\label{sec:verify}
This section showcases two simulated examples and two real data applications. The code is available at \url{https://github.com/zhedongliu/LGOCV}.\\ \\
We start with a simulation that tests the approximation accuracy in a multilevel model with various response types. Following this, a time series forecasting simulation is presented. This allows LGOCV results with automatically constructed groups to be compared to the LFOCV. We then delve into disease mapping, contrasting group constructions derived from various strategies. Finally, we apply our methodology to intricate models using a large dataset, as documented by \cite{lowe2021combined}. For the construction of the leave-out group for LGOCV, we used Algorithm 1 in Section \ref{sec:auto}. The procedures detailed in Sections \ref{sec:auto} and \ref{sec:method} have been integrated into R-INLA, ensuring that all computational tasks in this section are executed through R-INLA.

\subsubsection*{Simulated Multilevel Model with Various Responses}

This example is a simulation that demonstrates the accuracy of the approximation described in Section \ref{sec:method}. The main purpose is to compare $\pi(y_i|\boldsymbol{y}_{-I_i})$ computed using an approximation in Section \ref{sec:method} and the same quantity computed using MCMC. Furthermore, we use automatic group construction with the number of level sets equal to $1$, corresponding to predicting a data point from a new class.

We simulate data according to the following process. Initially, we simulate 10 class means, denoted as $\boldsymbol{s}$, from a standard normal distribution. Next, we compute $100$ linear predictors, $\eta_i = \mu + s_{j(i)}$, where $\mu = \log(10)$ and $j(i)$ is a function mapping data index $i$ to the group index $j$.  For this function, we set $j(i) = \lceil \frac{i}{10}\rceil$, where the ceiling function, $\lceil x \rceil$, rounds a number up to the nearest integer. We generate responses according to the linear predictor and one of three response types; Gaussian, Binomial and Exponential. The mean of the Gaussian response is $\eta_i$, and the standard deviation is 0.1. We generate binomial responses with a success probability of $\frac{1}{1+e^{-\eta_j}}$ for 20 trials. The exponential responses are generated with a mean of $e^{\eta_j}$.

We consider the model,
\begin{equation}
\label{eq:exp_group}
    \begin{split}
        &\log(\tau_s) \sim N(0,10^{-4}),   \quad  \mu \sim N(0,10^{-4}),   \quad  s_j|\tau_s \sim N(0,\tau_s), \\
        &\eta_i = \mu + s_j,  \quad  y_i|\eta_i \sim \text{response model}(\eta_i),
    \end{split}
\end{equation}
where the second parameter of the Gaussian distribution is the precision, and the likelihood is specified according to the data generation process with the given response model.

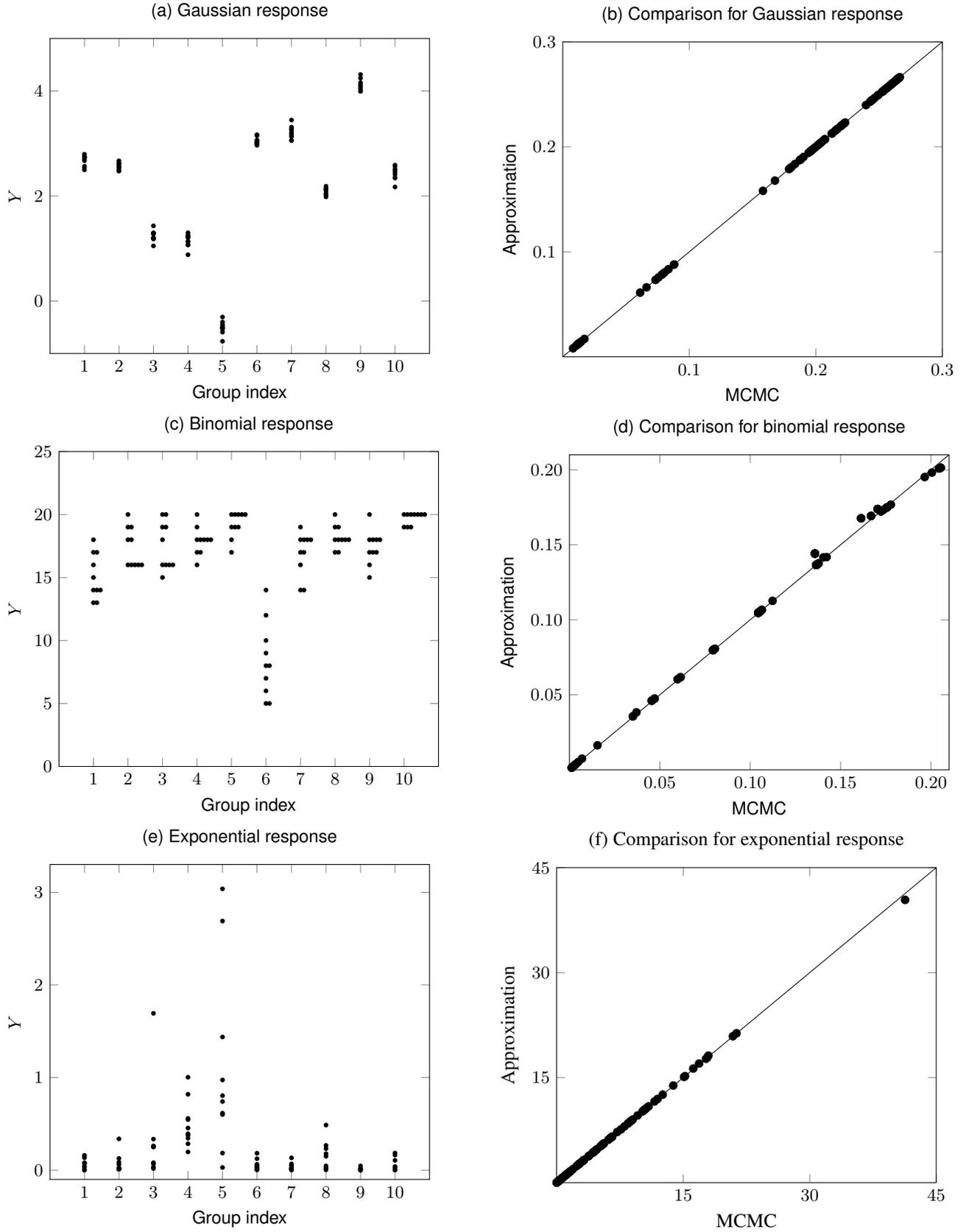
\begin{figure}[htp]
 \begin{subfigure}{0.5\textwidth}
  \begin{tikzpicture}[font=\sffamily\small]
  \begin{axis}[
    xlabel={Group index},
    ylabel={$Y$},
    xmin= 0 , xmax= 11,
    ymin= -1,  ymax= 5,
    xtick={1,2,3,4,5,6,7,8,9,10},
    grid style=dashed,
    title = (a) Gaussian response
  ]\addplot[
    only marks,
    mark size=1pt]
  coordinates {
    (1,2.757)(1,2.498)(1,2.739)(1,2.545)(1,2.673)(1,2.567)(1,2.740)(1,2.796)(1,2.730)(1,2.707)(2,2.625)(2,2.618)(2,2.553)(2,2.474)(2,2.620)(2,2.564)(2,2.528)(2,2.490)(2,2.671)(2,2.607)(3,1.294)(3,1.049)(3,1.182)(3,1.189)(3,1.296)(3,1.432)(3,1.203)(3,1.197)(3,1.295)(3,1.276)(4,1.216)(4,1.235)(4,1.299)(4,1.066)(4,1.132)(4,1.206)(4,1.135)(4,0.881)(4,1.074)(4,1.253)(5,-0.492)(5,-0.502)(5,-0.306)(5,-0.505)(5,-0.536)(5,-0.451)(5,-0.768)(5,-0.594)(5,-0.506)(5,-0.402)(6,3.150)(6,2.991)(6,3.001)(6,2.967)(6,3.042)(6,3.029)(6,3.168)(6,3.023)(6,2.999)(6,3.067)(7,3.311)(7,3.132)(7,3.148)(7,3.447)(7,3.057)(7,3.184)(7,3.202)(7,3.279)(7,3.254)(7,3.059)(8,2.029)(8,2.119)(8,2.099)(8,2.153)(8,2.126)(8,2.013)(8,2.143)(8,2.186)(8,2.046)(8,1.984)(9,4.043)(9,4.100)(9,4.314)(9,4.158)(9,4.076)(9,3.996)(9,4.246)(9,3.995)(9,4.095)(9,4.130)(10,2.589)(10,2.452)(10,2.342)(10,2.506)(10,2.353)(10,2.472)(10,2.567)(10,2.415)(10,2.504)(10,2.173)
  };
  \end{axis}
  \end{tikzpicture}
\end{subfigure}
\hspace{.5cm}
\begin{subfigure}{0.\textwidth}
  \begin{tikzpicture}[font=\sffamily\small]
    \begin{axis}[
    xlabel={MCMC},
    ylabel={Approximation},
    xmin= 0 , xmax= 0.3,
    ymin= 0,  ymax= 0.3,
    title = (b) Comparison for Gaussian response,
    xtick={.1,.2,.3,.4},
    ytick={.1,.2,.3,.4}
  ]\addplot[
    only marks,
    mark size=2pt]
  coordinates {
   (0.2428102,0.2428598)(0.2594905,0.2593678)(0.2442556,0.2442928)(0.2570534,0.2569634)(0.2491109,0.2490967)(0.2558300,0.2557525)(0.2441817,0.2442196)(0.2396937,0.2397585)(0.2448926,0.2449236)(0.2466143,0.2466273)(0.2523045,0.2522332)(0.2527623,0.2526797)(0.2565990,0.2564229)(0.2604690,0.2602325)(0.2526720,0.2525916)(0.2559913,0.2558291)(0.2579085,0.2577056)(0.2597749,0.2595444)(0.2492973,0.2492970)(0.2534536,0.2533537)(0.2069745,0.2070446)(0.1788330,0.1788453)(0.1942535,0.1942758)(0.1951046,0.1951303)(0.2072115,0.2072819)(0.2222649,0.2222251)(0.1966822,0.1967146)(0.1959852,0.1960146)(0.2070896,0.2071599)(0.2050288,0.2050951)(0.1976432,0.1976022)(0.1998010,0.1997661)(0.2071177,0.2071338)(0.1799014,0.1798880)(0.1877142,0.1876856)(0.1964447,0.1964021)(0.1881168,0.1880871)(0.1582450,0.1580945)(0.1808647,0.1808509)(0.2019534,0.2019287)(0.0126493,0.0126226)(0.0124644,0.0124393)(0.0170987,0.0170392)(0.0124012,0.0123767)(0.0117966,0.0117763)(0.0135335,0.0134982)(0.0081812,0.0081272)(0.0107400,0.0107213)(0.0123803,0.0123560)(0.0146373,0.0145914)(0.2034899,0.2035057)(0.2207240,0.2207296)(0.2196581,0.2196659)(0.2231752,0.2231777)(0.2152885,0.2153071)(0.2166477,0.2166630)(0.2015341,0.2015469)(0.2172898,0.2173035)(0.2198587,0.2198661)(0.2126159,0.2126391)(0.1836015,0.1837347)(0.2045625,0.2045869)(0.2027073,0.2027387)(0.1676922,0.1678230)(0.2129460,0.2129731)(0.1985136,0.1985676)(0.1964532,0.1965203)(0.1873360,0.1874557)(0.1903049,0.1904099)(0.2127522,0.2127786)(0.2636114,0.2634592)(0.2653142,0.2652256)(0.2650333,0.2649278)(0.2656702,0.2656146)(0.2653988,0.2653166)(0.2631729,0.2630110)(0.2655793,0.2655134)(0.2658519,0.2658266)(0.2640177,0.2638755)(0.2623302,0.2621511)(0.0834937,0.0834295)(0.0782565,0.0782123)(0.0611615,0.0611473)(0.0732602,0.0732347)(0.0804445,0.0803910)(0.0878847,0.0878143)(0.0662082,0.0661923)(0.0880424,0.0879720)(0.0787547,0.0787083)(0.0756373,0.0756038)(0.2546952,0.2545798)(0.2610590,0.2609783)(0.2644278,0.2643761)(0.2588315,0.2587336)(0.2641657,0.2641135)(0.2603017,0.2602143)(0.2558626,0.2557511)(0.2623777,0.2623104)(0.2589132,0.2588157)(0.2664362,0.2662942)
    };
    
    \addplot[]
  coordinates {
   (0,0)(0.5,0.5)
    };
  \end{axis}
  \end{tikzpicture}
\end{subfigure}
\\
\begin{subfigure}{0.5\textwidth}
  \begin{tikzpicture}[font=\sffamily\small]
  \begin{axis}[
    xlabel={Group index},
    ylabel={$Y$},
    xmin= 0 , xmax= 11,
    ymin= 0,  ymax= 25,
    xtick={1,2,3,4,5,6,7,8,9,10},
    grid style=dashed,
    title = (c) Binomial response
  ]\addplot[
    only marks,
    mark size=1pt]
  coordinates {
(1.0,13.0)(1.1,13.0)(1.0,14.0)(1.1,14.0)(1.2,14.0)(1.0,15.0)(1.0,16.0)(1.0,17.0)(1.1,17.0)(1.0,18.0)(2.0,16.0)(2.1,16.0)(2.2,16.0)(2.3,16.0)(2.4,16.0)(2.0,18.0)(2.1,18.0)(2.0,19.0)(2.1,19.0)(2.0,20.0)(3.0,15.0)(3.0,16.0)(3.1,16.0)(3.2,16.0)(3.3,16.0)(3.0,18.0)(3.0,19.0)(3.1,19.0)(3.0,20.0)(3.1,20.0)(4.0,16.0)(4.0,17.0)(4.1,17.0)(4.0,18.0)(4.1,18.0)(4.2,18.0)(4.3,18.0)(4.4,18.0)(4.0,19.0)(4.0,20.0)(5.0,17.0)(5.0,18.0)(5.0,19.0)(5.1,19.0)(5.2,19.0)(5.0,20.0)(5.1,20.0)(5.2,20.0)(5.3,20.0)(5.4,20.0)(6.0,5.0)(6.1,5.0)(6.0,6.0)(6.0,7.0)(6.0,8.0)(6.1,8.0)(6.0,9.0)(6.0,10.0)(6.0,12.0)(6.0,14.0)(7.0,14.0)(7.1,14.0)(7.0,16.0)(7.0,17.0)(7.1,17.0)(7.0,18.0)(7.1,18.0)(7.2,18.0)(7.3,18.0)(7.0,19.0)(8.0,17.0)(8.1,17.0)(8.0,18.0)(8.1,18.0)(8.2,18.0)(8.3,18.0)(8.4,18.0)(8.0,19.0)(8.1,19.0)(8.0,20.0)(9.0,15.0)(9.0,16.0)(9.0,17.0)(9.1,17.0)(9.2,17.0)(9.0,18.0)(9.1,18.0)(9.2,18.0)(9.3,18.0)(9.0,20.0)(10.0,19.0)(10.1,19.0)(10.2,19.0)(10.0,20.0)(10.1,20.0)(10.2,20.0)(10.3,20.0)(10.4,20.0)(10.5,20.0)(10.6,20.0)
  };
  \end{axis}
  \end{tikzpicture}
\end{subfigure}
\hspace{.5cm}
\begin{subfigure}{0.5\textwidth}
    \begin{tikzpicture}[font=\sffamily\small]
    \begin{axis}[
    xlabel={MCMC},
    ylabel={Approximation},
    xmin= 0 , xmax= 0.21,
    ymin= 0,  ymax= 0.21,
    title = (d) Comparison for binomial response,
    xtick={.05,.1,.15,0.2},
    ytick={.05,.1,.15,0.2},
    y tick label style={
        /pgf/number format/.cd,
        fixed,
        fixed zerofill,
        precision=2,
        /tikz/.cd
    },
    x tick label style={
        /pgf/number format/.cd,
        fixed,
        fixed zerofill,
        precision=2,
        /tikz/.cd
    }
  ]\addplot[
    only marks,
    mark size=2pt]
  coordinates {
(0.142279,0.141826)(0.035096,0.035605)(0.106430,0.106602)(0.106430,0.106602)(0.079484,0.079928)(0.045540,0.046083)(0.045540,0.046083)(0.045540,0.046083)(0.035096,0.035605)(0.059824,0.060362)(0.205346,0.201418)(0.079773,0.080032)(0.079773,0.080032)(0.079773,0.080032)(0.136929,0.136819)(0.175794,0.175058)(0.079773,0.080032)(0.175794,0.175058)(0.079773,0.080032)(0.136929,0.136819)(0.175390,0.174708)(0.079886,0.080139)(0.136814,0.136725)(0.061534,0.061807)(0.175390,0.174708)(0.079886,0.080139)(0.079886,0.080139)(0.204510,0.200903)(0.204510,0.200903)(0.079886,0.080139)(0.136513,0.136549)(0.104891,0.105083)(0.200537,0.198154)(0.136513,0.136549)(0.136513,0.136549)(0.173707,0.173320)(0.136513,0.136549)(0.136513,0.136549)(0.080513,0.080744)(0.104891,0.105083)(0.170538,0.173863)(0.166880,0.169282)(0.170538,0.173863)(0.166880,0.169282)(0.112373,0.112618)(0.170538,0.173863)(0.170538,0.173863)(0.140491,0.141646)(0.166880,0.169282)(0.170538,0.173863)(0.001256,0.001365)(0.004825,0.005169)(0.001735,0.001880)(0.015526,0.016326)(0.001256,0.001365)(0.003388,0.003645)(0.003388,0.003645)(0.002412,0.002605)(0.006988,0.007446)(0.037056,0.038308)(0.104493,0.104654)(0.137709,0.137504)(0.047119,0.047472)(0.104493,0.104654)(0.137709,0.137504)(0.177862,0.176808)(0.047119,0.047472)(0.079372,0.079685)(0.137709,0.137504)(0.137709,0.137504)(0.196542,0.195180)(0.136553,0.136700)(0.105512,0.105688)(0.172348,0.172297)(0.172348,0.172297)(0.105512,0.105688)(0.136553,0.136700)(0.136553,0.136700)(0.136553,0.136700)(0.136553,0.136700)(0.136928,0.136819)(0.136928,0.136819)(0.205485,0.201418)(0.061362,0.061679)(0.104454,0.104645)(0.136928,0.136819)(0.104454,0.104645)(0.104454,0.104645)(0.136928,0.136819)(0.079734,0.080032)(0.161384,0.167712)(0.135811,0.144175)(0.161384,0.167712)(0.161384,0.167712)(0.135811,0.144175)(0.135811,0.144175)(0.135811,0.144175)(0.135811,0.144175)(0.135811,0.144175)(0.135811,0.144175)
    };
    
    \addplot[]
  coordinates {
   (0,0)(0.21,0.21)
    };
  \end{axis}
  \end{tikzpicture}
\end{subfigure}
\\
\begin{subfigure}{0.5\textwidth}
  \begin{tikzpicture}[font=\sffamily\small]
  \begin{axis}[
    xlabel={Group index},
    ylabel={$Y$},
    xmin= 0 , xmax= 11,
    ymin= -0.1,  ymax= 3.3,
    xtick={1,2,3,4,5,6,7,8,9,10},
    grid style=dashed,
    title = (e) Exponential response
  ]\addplot[
    only marks,
    mark size=1pt]
  coordinates {
    (1,0.134)(1,0.160)(1,0.044)(1,0.008)(1,0.076)(1,0.034)(1,0.074)(1,0.152)(1,0.000)(1,0.080)(2,0.338)(2,0.011)(2,0.012)(2,0.018)(2,0.011)(2,0.087)(2,0.127)(2,0.012)(2,0.063)(2,0.027)(3,1.693)(3,0.254)(3,0.251)(3,0.335)(3,0.035)(3,0.069)(3,0.081)(3,0.018)(3,0.262)(3,0.253)(4,1.003)(4,0.197)(4,0.819)(4,0.285)(4,0.393)(4,0.378)(4,0.546)(4,0.455)(4,0.558)(4,0.345)(5,0.974)(5,0.029)(5,3.037)(5,0.804)(5,2.690)(5,0.615)(5,1.438)(5,0.741)(5,0.185)(5,0.600)(6,0.028)(6,0.005)(6,0.033)(6,0.005)(6,0.183)(6,0.124)(6,0.063)(6,0.041)(6,0.051)(6,0.008)(7,0.133)(7,0.014)(7,0.026)(7,0.008)(7,0.011)(7,0.001)(7,0.025)(7,0.046)(7,0.053)(7,0.068)(8,0.017)(8,0.487)(8,0.004)(8,0.033)(8,0.150)(8,0.046)(8,0.233)(8,0.267)(8,0.178)(8,0.018)(9,0.001)(9,0.007)(9,0.008)(9,0.024)(9,0.017)(9,0.002)(9,0.014)(9,0.000)(9,0.001)(9,0.046)(10,0.035)(10,0.039)(10,0.185)(10,0.002)(10,0.033)(10,0.003)(10,0.013)(10,0.025)(10,0.163)(10,0.106)
  };
  \end{axis}
  \end{tikzpicture}
\end{subfigure}
\hspace{.5cm}
\begin{subfigure}{0.5\textwidth}
    \begin{tikzpicture}
    \begin{axis}[
    xlabel={MCMC},
    ylabel={Approximation},
    xmin= 0 , xmax= 45,
    ymin= 0,  ymax= 45,
    title = (f) Comparison for exponential response,
    xtick={15,30,45},
    ytick={15,30,45}
  ]\addplot[
    only marks,
    mark size=2pt]
  coordinates {
(1.4844855,1.4847962)(1.2158650,1.2158838)(4.4110752,4.4126623)(12.5740737,12.5708873)(2.6919659,2.6929153)(5.3501086,5.3516814)(2.7759974,2.7769693)(1.2828633,1.2829661)(41.3104466,40.39796)(2.5823559,2.5832733)(0.4519270,0.4473409)(10.4972171,10.4970017)(10.4176980,10.4174756)(8.0348820,8.0343744)(10.8856899,10.8855769)(2.3671347,2.3670163)(1.5932725,1.5930218)(10.3718205,10.3715963)(3.2451947,3.2449639)(6.3428877,6.3422486)(0.0242785,0.0238240)(0.6400406,0.6402698)(0.6522135,0.6524825)(0.4256592,0.4252925)(5.5755335,5.5925577)(3.0821893,3.0922256)(2.6223067,2.6310122)(8.7450430,8.7649885)(0.6132481,0.6133910)(0.6453898,0.6456364)(0.0608752,0.0602266)(0.9158815,0.9168842)(0.0894671,0.0886800)(0.5376937,0.5371846)(0.3244550,0.3235454)(0.3457287,0.3448538)(0.1861406,0.1851392)(0.2545086,0.2535136)(0.1791116,0.1781158)(0.3993867,0.3986072)(0.0492330,0.0480268)(7.1815428,7.2287540)(0.0037533,0.0035808)(0.0739122,0.0723667)(0.0049821,0.0047642)(0.1281061,0.1260820)(0.0209091,0.0202382)(0.0875301,0.0858321)(1.0282230,1.0297399)(0.1344924,0.1324329)(6.1376227,6.1344641)(15.2192079,15.2164360)(5.4159037,5.4129136)(15.1055862,15.1027703)(1.0459178,1.0426777)(1.6459137,1.6443694)(3.2158975,3.2141281)(4.6668157,4.6641098)(3.8737048,3.8715080)(11.9530860,11.9485627)(1.5564815,1.5497703)(9.0051377,8.9993539)(6.4089309,6.4028409)(11.5960754,11.5929508)(10.1779560,10.1727760)(20.8925514,20.9157624)(6.5142311,6.5081245)(4.2677097,4.2629764)(3.7571338,3.7528163)(3.0604729,3.0566151)(8.7685039,8.7765389)(0.2492016,0.2485218)(17.7096636,17.7022394)(5.6045690,5.6126105)(1.2877839,1.2892058)(4.3009283,4.3066543)(0.7369717,0.7373672)(0.6093730,0.6095288)(1.0471436,1.0481205)(8.4523908,8.4604906)(16.8858245,17.0103505)(10.6525430,10.6532526)(10.2332306,10.2297077)(6.5413911,6.5155031)(7.6271896,7.6063250)(13.8207624,13.8692192)(8.4484863,8.4317158)(17.9881025,18.1498787)(16.1972237,16.3031942)(4.4213709,4.3658819)(5.1853435,5.1832837)(4.7890892,4.7870870)(1.0247171,1.0227632)(21.3223328,21.3353537)(5.4275974,5.4255066)(17.8479066,17.8540018)(9.5989082,9.5962056)(6.5856971,6.5830639)(1.2011603,1.1997269)(1.9390296,1.9381801)
    };
    \addplot[]
  coordinates {
   (0,0)(50,50)
    };
  \end{axis}
  \end{tikzpicture}
\end{subfigure}
\caption{Comparison of $\pi(y_i|\boldsymbol{y}_{-I_i})$ from approximations and MCMC. First column: $y$-axis shows response value, $x$-axis shows group index. Second column: $y$-axis shows LGOCV from proposed approximation, $x$-axis shows LGOCV from MCMC.}
\label{simulated_data}
\end{figure}

As a reference, we let the MCMC runs for $10^8$ iterations, which makes the Monte Carlo errors negligible. The large size of MCMC samples is required because the predictive distributions are influenced by the tails of $\pi(\eta_i|\boldsymbol{\theta},\boldsymbol{y}_{-I_i})$. In Figure \ref{simulated_data}, (a), (c), and (d) show the data against its group index, which presents a clear group structure; (b), (d), and (e) show the comparison of $\pi(y_i|\boldsymbol{y}_{-I_i})$ obtained from the approximations and MCMC. We use Rstan \cite{Rstan} for the MCMC.

This example shows that the approximations are highly accurate. When the response is Gaussian, the approximation almost equals the MCMC results, where the main difference is due to MCMC sampling errors, as our approach is exact up to numerical integration in this case. Also, under both non-Gaussian cases, the results are close to the long-run MCMC results. 


\subsubsection*{Time Series Forecasting}
In this example, we will demonstrate how the automatic LGOCV method can measure the forecasting performance of a time series model, while LOOCV is not effective in doing so. 

We will first simulate $2000$ data points using the following procedure: We will simulate an AR(1) time series by using $u_i = 0.9u_{i-1} + \epsilon_{u_i}$, where $\epsilon_{u_i}$ follows a standard Gaussian distribution. Next, we will compute linear predictors by calculating $\eta_i = \mu + u_i$, with $\mu$ set to 2. Finally, the Gaussian responses have mean $\eta_i$, and a standard deviation of $0.1$.

We fit a time series model on the simulated data:
\begin{equation*}
	\label{eq:time_simulated}
	\begin{split}
	& \mu \sim N(0,10^{-4}), \quad \boldsymbol{u} \sim N(0,\boldsymbol{Q}_u),\\
        & \eta_i = \mu + u_i,  \quad y_i|\eta_i \sim N(\eta_i, 100),
	\end{split}
\end{equation*}
where $\boldsymbol{Q}_u$ is determined by an AR(1) model with the true parameters.

The prediction task is $k$ steps forward forecasting for $k = \{1,2,\dots , 10 \}$ using the true model. The natural cross-validation for these prediction tasks is LFOCV. To replicate the LFOCV, the group in LGOCV for testing point $y_i$ and $k$ steps forward prediction includes $\boldsymbol{y}_{(i-k+1):n}$. We can compute LFOCV for every $k$, denoted by LFOCV($k$). To make the training set similar to the data set, the last $500$ data points will be used as testing points, which means $i = \{1501,\dots 2000\}$ in (\ref{u_LGOCV}), and the quantity is averaged over $500$ data points. We can also compute LGOCV using automatically constructed groups with the number of level sets, $m = \{1,2,\dots, 10 \}$, denoted by LGOCV($m$). In this setting, the automatically constructed group for a testing point $y_i$ with a number of level sets equal to $m$ includes $\boldsymbol{y}_{\max(1,i-m+1):\min(n,i+m-1)}$. Also, LGOCV($1$) is equivalent to LOOCV in this model.

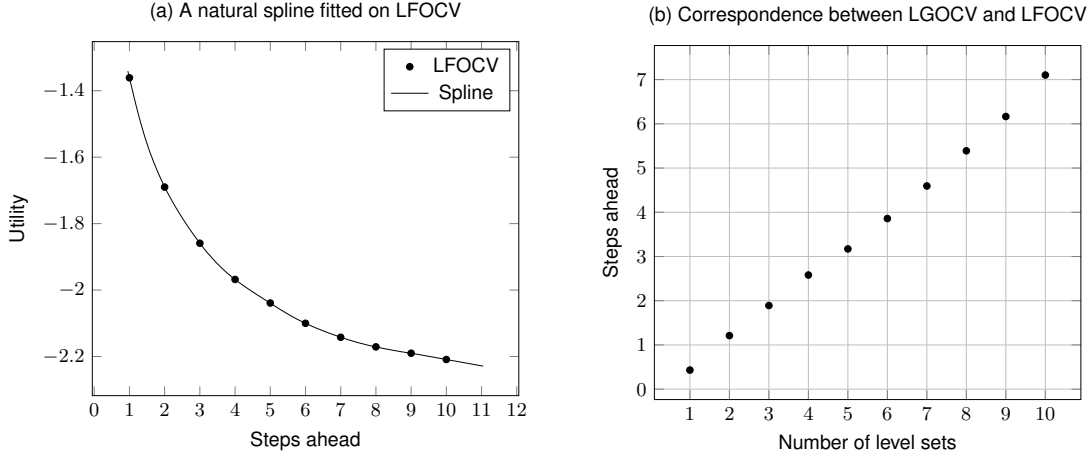
\begin{figure}
\resizebox{0.9\textwidth}{!}{
\begin{subfigure}{.5\textwidth}
\begin{tikzpicture}[font=\sffamily\small]
          \begin{axis}
             [
               title = (a) A natural spline fitted on LFOCV 
              ,xlabel = Steps ahead
              ,ylabel = Utility
              ,ymajorticks = true
              ,xtick = {0,1,2,3,4,5,6,7,8,9,10,11,12}
             ]
             \addplot+[only marks, color = black,mark options={fill = black}, mark size=1.5pt] coordinates
             {(1.000,-1.361)(2.000,-1.690)(3.000,-1.859)(4.000,-1.968)(5.000,-2.039)(6.000,-2.100)(7.000,-2.142)(8.000,-2.171)(9.000,-2.190)(10.000,-2.209)};
             \addplot+[color = black,mark = none] coordinates
             {(11.049,-2.229)(10.579,-2.220)(10.108,-2.211)(9.637,-2.202)(9.163,-2.193)(8.684,-2.185)(8.222,-2.176)(7.818,-2.167)(7.484,-2.158)(7.199,-2.149)(6.945,-2.140)(6.707,-2.131)(6.483,-2.122)(6.273,-2.113)(6.079,-2.104)(5.899,-2.095)(5.732,-2.086)(5.577,-2.077)(5.431,-2.068)(5.291,-2.059)(5.156,-2.050)(5.023,-2.041)(4.890,-2.032)(4.756,-2.023)(4.624,-2.014)(4.494,-2.005)(4.367,-1.996)(4.243,-1.987)(4.124,-1.978)(4.010,-1.969)(3.903,-1.960)(3.801,-1.951)(3.705,-1.942)(3.614,-1.933)(3.528,-1.924)(3.446,-1.915)(3.369,-1.906)(3.294,-1.897)(3.223,-1.889)(3.154,-1.880)(3.087,-1.871)(3.022,-1.862)(2.958,-1.853)(2.896,-1.844)(2.834,-1.835)(2.774,-1.826)(2.715,-1.817)(2.656,-1.808)(2.600,-1.799)(2.544,-1.790)(2.489,-1.781)(2.436,-1.772)(2.383,-1.763)(2.332,-1.754)(2.282,-1.745)(2.233,-1.736)(2.186,-1.727)(2.139,-1.718)(2.094,-1.709)(2.050,-1.700)(2.007,-1.691)(1.966,-1.682)(1.925,-1.673)(1.886,-1.664)(1.848,-1.655)(1.811,-1.646)(1.775,-1.637)(1.740,-1.628)(1.705,-1.619)(1.672,-1.610)(1.640,-1.601)(1.609,-1.593)(1.578,-1.584)(1.549,-1.575)(1.520,-1.566)(1.492,-1.557)(1.464,-1.548)(1.437,-1.539)(1.411,-1.530)(1.386,-1.521)(1.361,-1.512)(1.337,-1.503)(1.313,-1.494)(1.289,-1.485)(1.267,-1.476)(1.244,-1.467)(1.222,-1.458)(1.200,-1.449)(1.179,-1.440)(1.158,-1.431)(1.137,-1.422)(1.116,-1.413)(1.096,-1.404)(1.076,-1.395)(1.055,-1.386)(1.035,-1.377)(1.015,-1.368)(0.995,-1.359)(0.975,-1.350)(0.955,-1.341)};
             \addlegendentry{LFOCV} 
			 \addlegendentry{Spline}           
            \end{axis}
\end{tikzpicture}
\end{subfigure}
\hspace{1.1cm}
\begin{subfigure}{.5\textwidth}
\begin{tikzpicture}[font=\sffamily\small]
          \begin{axis}
             [
               title = (b) Correspondence between LGOCV and LFOCV
              ,xlabel = Number of level sets
              ,ylabel = Steps ahead
              ,ymajorticks = true
              ,grid=both
              ,xtick= {0,1,2,3,4,5,6,7,8,9,10}
              ,ytick = {0,1,2,3,4,5,6,7,8,9,10}
             ]
             \addplot+[only marks, color = black,mark options={fill = black}, mark size=1.5pt] coordinates
             {(1.000,0.432)(2.000,1.211)(3.000,1.891)(4.000,2.581)(5.000,3.171)(6.000,3.858)(7.000,4.594)(8.000,5.390)(9.000,6.165)(10.000,7.103)};
             \addplot+[color = black,mark = none] coordinates
             {(1,1)(10,10)};
            \end{axis}
\end{tikzpicture}
\end{subfigure}
}
\medskip
\caption{Comparison of Automatic LGOCV and LFOCV. LOOCV measures approximately 0.4 steps forward forecasting. LGOCV(2) roughly represents a one-step forward forecasting performance.}
\label{fig:time series simulation}
\end{figure}

To compare LGOCV and LFOCV, we will fit a natural spline to have LFOCV(t) for $t$ as a real number (see Figure \ref{fig:time series simulation} (a)) and map the number of level sets in LGOCV to the steps ahead in LFOCV (see Figure \ref{fig:time series simulation} (b)). We can see that LOOCV measures approximately 0.4 steps forward forecasting when the simplest prediction task is one step forward forecasting. LGOCV(2) represents roughly a one-step forward forecasting performance of the model. As the number of level sets increases in LGOCV, it represents more steps forward forecasting performance. Note that the specific translation between the automatic LGOCV and LFOCV is only valid in this model and may not be applicable in other models.

\subsubsection*{Disease Mapping}
In this example, we will present groups constructed by different automatic group construction strategies. We will see the differences between those groups and get an idea to choose a proper group construction strategy. 

We applied a disease mapping model to data detailing cancer incidence by location \cite{besag1991bayesian,wakefield2000bayesian,held2005towards}. This dataset captures oral cavity cancer cases in Germany from 1986-1990 \cite{held2005towards}. The response $y_i$ indicates the cases in area $i$ over five years. The case count in each region is influenced by its population and age distribution. The expected case count $E_i$ in the region $i$ is derived from its age distribution and population, ensuring $\sum_i y_i = \sum_i E_i$. Additionally, the covariate $x_i$ represents tobacco consumption in area $i$.

We fit the following model on the data set:
\begin{equation}
	\begin{split}
	&y_i|\eta_i \sim \text{Poisson}(E_i\exp{(\eta_i)}) \\
	&\eta_i = \mu + f_\text{rw}(x_i) + u_i + v_i,\\
	\end{split}
\end{equation}
where $\mu$ is an intercept, $\boldsymbol{u}$ is a spatially structured component, $\boldsymbol{v}$ is an unstructured component \cite{krainski2018advanced}, and $\boldsymbol{f}_{rw}$ is an intrinsic second-order random-walk model of the covariate $x_i$ \cite{rue2005gaussian}.

\begin{figure}[h!]
\centering
\includegraphics[width=0.81\textwidth]{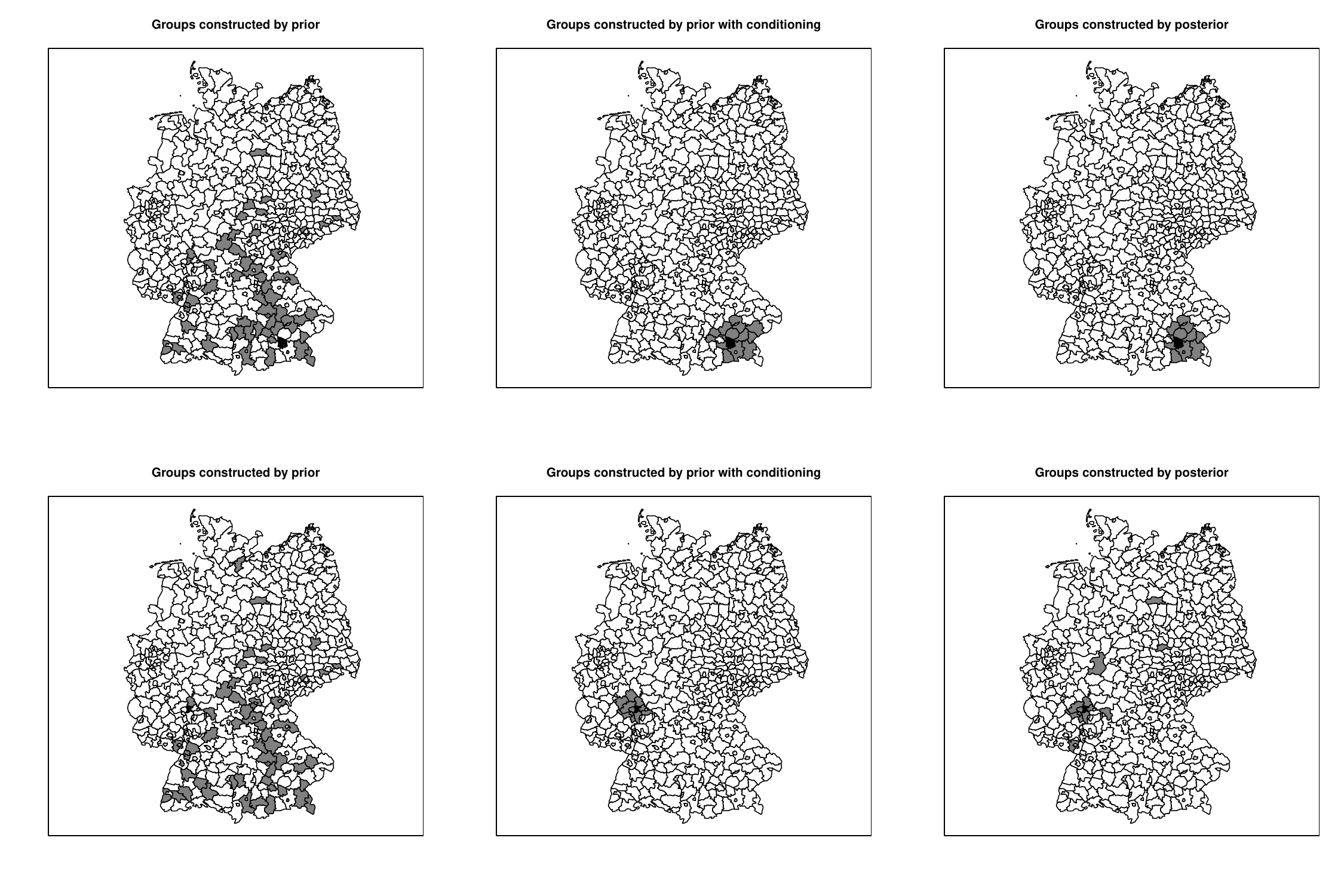}
\caption{Groups by different automatic construction strategies. The testing point is located in the black region, and the data in the group are located in the grey regions. In (d), the group constructed by posteriors contains some non-spatial patterns.}
\label{fig:germany}
\end{figure}

In Figure \ref{fig:germany}, we illustrate groups formed through various automatic group construction strategies. The testing point is located in the black region, while the data in the group are located in grey areas. As seen in Figure \ref{fig:germany} (a) and (b), Groups from $\boldsymbol{R}_{\text{prior}}$ focus solely on spatial effects. Groups from $\boldsymbol{R}_{\text{post}}$ exhibit mostly strong spatial patterns, such as Figure \ref{fig:germany} (c). Yet, some points, like in Figure \ref{fig:germany} (d), indicate non-spatial patterns. This arises as all model components, including fixed and random effects, priors, and the response variable, are considered.

The spatial patterns in posterior groups may justify incorporating spatial effects into the model, given that data retains this pattern in correlation. In practice, groups from $\boldsymbol{R}_{\text{post}}$ offer a more balanced representation. However, $\boldsymbol{R}_{\text{prior}}$ with selective effects resemble those from manually defined groups.

\subsubsection*{Dengue Risk in Brazil}
\label{Dengue}
In this real-world example, we will demonstrate the scalability and adaptability of the automatic LGOCV method in a complex model structure and a large sample$~$size. The automatically constructed model-based groups are consistent with the domain knowledge that dengue disease is prevalent in summer. 

We will repeat the variable selection process as shown in \cite{lowe2021combined} using the automatic LGOCV. The model chosen by LGOCV is considered to have better predictive power for longer-range predictions, than those selected based on other criteria because the most informative data points for predicting the target are excluded from the training set.

The models study the influence of extreme hydrometeorological hazards on dengue risk, factoring in Brazil's urbanization levels. Our dataset, with $127,224$ samples representing $12,895,293$ dengue cases, covers Brazil's 558 microregions from January 2001 to December 2019. Given the dataset's magnitude and the model's intricacy, LGOCV or LOOCV calculations require the approximation method detailed in Section \ref{sec:method}.

Data points include month, year, microregion, and state. The candidate covariates encompass the monthly average of daily minimum ($T_{min}$) and maximum temperatures ($T_{max}$), the palmer drought severity index (PDSI), the urbanization levels: overall ($u$), centered at high ($u_1$), intermediate ($u_2$), and more rural levels ($u_3$) and the access to water supply: overall ($w$) and centered at high-frequency shortages ($w_1$), intermediate ($w_2$), and low-frequency shortages ($w_3$). To preprocess these covariates' specifics, refer to \cite{rachel_lowe_2021_4632205}.

The data generating model is chosen to be negative binomial, to account for overdispersion. The latent field consists of a temporal component describing a state-specific seasonality using a cyclic first difference prior distribution and a spatial component describing year-specific spatially unstructured and structured random effects using a modified Besag-York-Mollie (BYM2) model with a scaled spatial component \cite{riebler2016intuitive}. The temporal component has replications for each state, and the spatial component has replications for each year. We can express the base model using the INLA-style formula,
\begin{equation*}
\begin{split}
	\text{y} \sim 1 &+ \text{covariates} + \text{f(month, model = ``rw1", replicate = state, cyclic = TRUE)}\\
	&+ \text{f(microregion, model = ``bym2", replicate = year)}.
\end{split}
\end{equation*}
In short, we write this model as $\text{y} \sim 1 + \text{covariates} + f_t + f_s.$
The number of parameters in this model is $21,567$ with $127,224$ observations for the full model. The appendix of \cite{lowe2021combined} and its repository \cite{rachel_lowe_2021_4632205} provide full details about the models and data. 

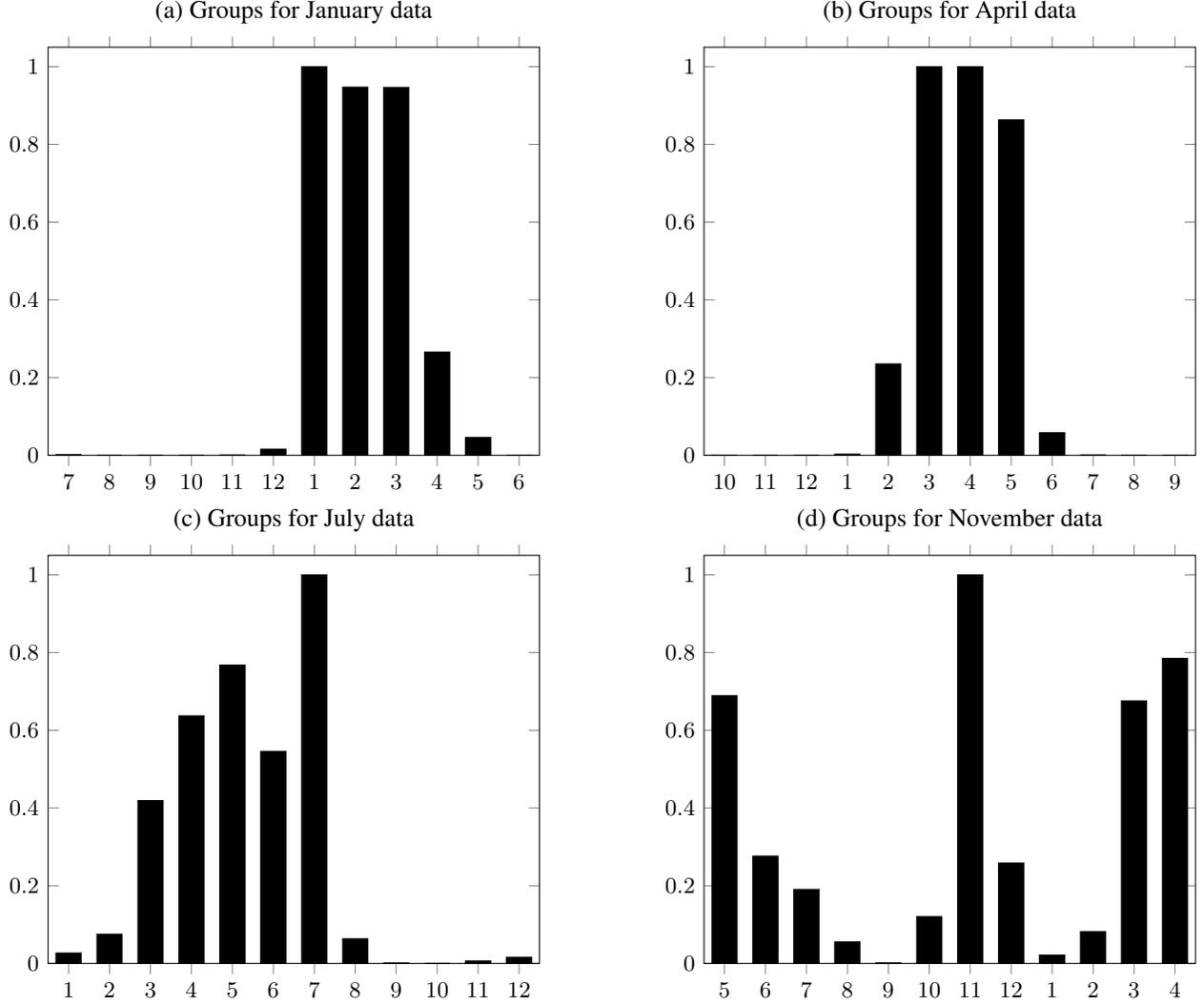
\begin{figure}[h!]

\begin{subfigure}[t]{0.55\textwidth}
\begin{tikzpicture}
\begin{axis} [ybar,xmin= .5 , xmax= 12.5, title = {(a) Groups for January data},
    ymin= 0,  ymax= 1.05,
    xtick = {1,2,3,4,5,6,7,8,9,10,11,12},
    xticklabels={$7$, $8$, $9$, $10$, $11$, $12$, $1$, $2$, $3$, $4$, $5$,$6$}]
\addplot[fill=black] coordinates {
    (1,0.00123)
    (2,0)
    (3,0)
    (4,0)
    (5,0.00009)
    (6,0.01556)
    (7,1)
    (8,0.94727)
    (9,0.94671)
    (10,0.26580)
    (11,0.04593)
    (12,0)
};
\end{axis}
\end{tikzpicture}
\end{subfigure}
\hfill
\begin{subfigure}[t]{0.55\textwidth}
\begin{tikzpicture}
\begin{axis} [ybar,xmin= .5 , xmax= 12.5,title = {(b) Groups for April data},
    ymin= 0,  ymax= 1.05,
    xtick = {1,2,3,4,5,6,7,8,9,10,11,12},
    xticklabels={ $10$, $11$, $12$, $1$, $2$, $3$, $4$, $5$,$6$,$7$, $8$, $9$,}]
\addplot[fill=black] coordinates {
    (1,0)
    (2,0)
    (3,0)
    (4,0.00302)
    (5,0.23505)
    (6,1)
    (7,1)
    (8,0.86295)
    (9,0.05810)
    (10,0.00019)
    (11,0)
    (12,0)
};
\end{axis}
\end{tikzpicture}
\end{subfigure}

\begin{subfigure}[t]{0.55\textwidth}
\begin{tikzpicture}
\begin{axis} [ybar,xmin= .5 , xmax= 12.5,title = {(c) Groups for July data},
    ymin= 0,  ymax= 1.05,
    xtick = {1,2,3,4,5,6,7,8,9,10,11,12},
    xticklabels={$1$, $2$, $3$, $4$, $5$,$6$,$7$, $8$, $9$, $10$, $11$, $12$}]
\addplot[fill=black] coordinates {
    (1,0.02688)
    (2,0.07508)
    (3,0.41898)
    (4,0.63714)
    (5,0.76769)
    (6,0.54527)
    (7,1)
    (8,0.06357)
    (9,0.00066)
    (10,0)
    (11,0.00623)
    (12,0.01566)
};
\end{axis}
\end{tikzpicture}
\end{subfigure}
\begin{subfigure}[t]{0.55\textwidth}
\begin{tikzpicture}
\begin{axis} [ybar,xmin= .5 , xmax= 12.5,title = {(d) Groups for November data},
    ymin= 0,  ymax= 1.05,
    xtick = {1,2,3,4,5,6,7,8,9,10,11,12},
    xticklabels={$5$,$6$,$7$, $8$, $9$, $10$, $11$, $12$, $1$, $2$, $3$, $4$}]
\addplot[fill=black] coordinates {
    (1,0.68893)
    (2,0.27627)
    (3,0.19053)
    (4,0.05527)
    (5,0.00085)
    (6,0.12035)
    (7,1)
    (8,0.25863)
    (9,0.02141)
    (10,0.08159)
    (11,0.67572)
    (12,0.78513)
};
\end{axis}
\end{tikzpicture}
\end{subfigure}

\medskip
\small
\caption{Groups for testing points from a specific month. $y$-axis: relative frequency, $x$-axis: month of data measurement in groups. The first half-year data are more informative for prediction. As shown in (c) and (d), even in July and November, the group often includes data consistent with the known summer prevalence of dengue. Note that dengue is prevalent in the summer months which are approximately November to February.}
\label{group_july_jan}
\end{figure}

The model accounts for temporal effects with spatial replicates and spatial effects with temporal replicates, complicated by various constraints. Given its intricacy and the lack of a clear prediction task, crafting groups for LGOCV manually is challenging. Hence, utilizing our automatic group construction through posterior correlation is beneficial. For model comparisons, using the same groups across different models is recommended. The base model, which only incorporates structured components, is chosen for group building. Most automatic groups cluster data from the same year, location, and nearby months to the testing points. Figure \ref{group_july_jan} displays the relative month frequencies in the group, given the testing points correspond to a specific month. The chart suggests the first half-year data better informs predictions. Even in July and November testing points, the group frequently includes that data, aligning with the known prevalence of dengue during summer. See Figure \ref{group_july_jan} (c) and (d) for details.

The results of model selection using deviance information criterion (DIC), LOOCV, and LGOCV ($m=2,3,4$) are presented in Table \ref{selection_res}. The candidate models are those referenced in \cite{rachel_lowe_2021_4632205}. To transform equation~\ref{u_LGOCV} into a loss function, we calculated its negative value. \\
From Table \ref{selection_res} we note that LOOCV prefers the spatio-temporal model that incorporates access to water while the spatio-temporal model with urbanization as a covariate, is preferred by LGOCV. This result is interesting since we can conclude that the same model might not necessarily perform well for short- and longer-range prediction. The practitioner thus needs to decide what the goal of the modeling is, and then choose the model to be used accordingly. If we want to predict dengue risk for a new unobserved area or time point, it seems that urbanization has a better prediction ability than access to water. Note also that as we increase the number of level sets, we are defining a prediction target with an increased range, thus moving further away from LOOCV.

\begin{table}[h!]
\centering
\begin{tabular}{|l|l|l|l|l|l|l|}
\hline
\multirow{2}{*}{Index} & \multirow{2}{*}{Model} & \multirow{2}{*}{DIC}  & \multirow{2}{*}{LOOCV} & \multicolumn{3}{c|}{LGOCV}   \\ \cline{5-7}
& & & & ($m=2$) & ($m=3$) & ($m=4$) \\ \hline
1 & $\text{y} \sim 1  + f_t + f_s$                                         & 3615.38   & 0.0151 & 0.0158 &  0.0206 & 0.0270 \\ \hline
2 &$\text{y} \sim 1 + T_{min} + f_t + f_s$                                & 1562.96    & 0.0064 & 0.0067 &  0.0088 & 0.0098 \\ \hline
3 &$\text{y} \sim 1 + T_{max} + f_t + f_s$                                & 2228.73   & 0.0091 & 0.0098 & 0.0133 & 0.0163 \\ \hline
4 &$\text{y} \sim 1 + PDSI + f_t + f_s$                                   & 2167.12   & 0.0092 & 0.0095 & 0.0126 &  0.0184 \\ \hline
5 &$\text{y} \sim 1 + PDSI + T_{min} + f_t + f_s$                         & 160.43  & 0.0006 & 0.0006 & 0.0012 & 0.0023 \\ \hline
6 &$\text{y} \sim 1 + PDSI + T_{max} + f_t + f_s$                         & 900.65  & 0.0038 & 0.0038 & 0.0057 & 0.0084 \\ \hline
7 &$\text{y} \sim 1 + PDSI + T_{min} + PDSI*u_1 +u + f_t + f_s$         & 38.21  & 0.0002 & 0*   & 0* & 0*  \\ \hline
8 &$\text{y} \sim 1 + PDSI + T_{min} + PDSI*u_2 +u + f_t + f_s$         & 39.13  & 0.0002 & 0* & 0* & 0* \\ \hline
9 &$\text{y} \sim 1 + PDSI + T_{min} + PDSI*u_3 +u + f_t + f_s$         & 28.64  & 0.0002 & 0* & 0* & 0* \\ \hline
10 &$\text{y} \sim 1 + PDSI + T_{min} + PDSI*w_1 +w + f_t + f_s$         & 6.68  &  0* & 0.0005  & 0* & 0.0014\\ \hline
11 &$\text{y} \sim 1 + PDSI + T_{min} + PDSI*w_2 +w + f_t + f_s$         & 0*  & 0* &  0.0005 & 0* & 0.0015\\ \hline
12 &$\text{y} \sim 1 + PDSI + T_{min} + PDSI*w_3 +w + f_t + f_s$         & 4.55 & 0*  & 0.0006 & 0* & 0.0014\\ \hline
\end{tabular}

\caption{Comparative evaluation of models for predicting variable \textit{y} based on various environmental factors. This table presents the model selection results, including each model's Deviance Information Criterion (DIC), LOOCV, and LGOCV scores.\\
Note: We offset DIC by 826841.66, LOOCV by 3.2721, LGOCV ($m=2$) by  3.314, LGOCV ($m=3$) by  3.3763 and LGOCV ($m=4$) by 3.4372.}
\label{selection_res}
\end{table}

\section{Discussion}
\label{sec:conc}

An over-reliance on LOOCV to evaluate predictive capacity in general persists in statistical practice, despite concerns raised in studies such as \cite{roberts2017cross,vehtari2019limitations}. LOOCV can provide an evaluation of short-range predictive ability with well-established asymptotics for some models. On the other hand, what can we do to evaluate the longer-range predictive ability of complex models? Various approaches for specific models, such as time series or spatial models have been proposed, where custom CV procedures are designed to mimic a longer-range prediction task than that of LOOCV. We have introduced an automated approach for evaluating the longer-range prediction ability of any latent Gaussian model, namely LGOCV. LGOCV is designed to be applicable to all models that are latent Gaussian models, and thus provides a framework for general longer-range predictive ability evaluation without the need for case-by-case considerations. Moreover, we propose a computationally efficient approach to calculate LGOCV scores and metrics based on the INLA methodology. We have shown that our approximate LGOCV implementation is almost exact when compared with the results from MCMC, albeit at a much lower computational cost. This enables practitioners to use the LGOCV approach for complex models and large data.\\ \\
Our approach is designed for latent Gaussian models and some ideas can thus be extended to the case of non-latent Gausian models with careful consideration of the computational cost associated with this endeavor. LGOCV for LGMs is computationally efficient in INLA, since it is fully parallelizable by computing the necessary quantities only at the mode of the hyperparameters. For huge data ($n > 10^6$) however, the cost will be high since the cost increases linearly in $n$, albeit much lower than other available approaches. For huge data, performing CV on a subset or constructing the groups manually could be considered. Nonetheless, for LGMs, the proposed LGOCV could be considered as the most feasible approach for longer-range predictive ability evaluation. \\ \\
The choice of the number of level sets determine the prediction task and thus the degree of independence between the leave-out group and the rest of the data. There is not a one-to-one correspondence between the number of level sets and the number of points to leave out as shown in the simulations and applications, although a higher number of level sets would imply a longer range for the prediction task, than a lower number. The choice of the number of level sets remains arbitrary since it is a user-defined parameter, we recommend a low number like $m=2$ or $m=3$ if there is no clear indication of what else $m$ should be. There exists no optimal value of $m$ in general, since it would imply different prediction tasks for different levels of dependency. In our applications, and those of others who have applied the LGOCV framework, it is shown that LGOCV provides the information we need to evaluate longer-range prediction ability, and complements the information from LOOCV. \\
It is pertinent to note that the proposed LGOCV do not replace a custom CV strategy designed by modelers, tailored for specific applications. We pose it as an alternative default strategy for longer-range prediction ability evaluation, that complements LOOCV in assessing the predictive ability of an LGM, while being computationally efficient and practical for real-world scenarios.

\section*{Acknowledgments}
The authors thank D.~Castro-Camilo, D.~Rustand, and E.~Krainski for valuable discussions and suggestions.

\appendix

\section[On the computation of Sigma and mu for eta]{On the computation of $\boldsymbol{\Sigma}_{\boldsymbol{\eta}_{I_i}}(\boldsymbol{\theta},\boldsymbol{y}_{-I_i})$ and $\boldsymbol{\mu}_{\boldsymbol{\eta}_{I_i}}(\boldsymbol{\theta},\boldsymbol{y}_{-I_i})$}

In this section, we let $I_i$ be $I$ and drop $\boldsymbol{\theta}$ to simplify the notation.
We have a random vector $\boldsymbol{\eta}_{I}|\boldsymbol{y} \sim N(\boldsymbol{\mu}_{\boldsymbol{\eta}_I}(\boldsymbol{y}), \boldsymbol{\Sigma}_{\boldsymbol{\eta}_I}(\boldsymbol{y}))$, which can be viewed as a posterior distribution with prior $\boldsymbol{\eta}_{I}|\boldsymbol{y}_{-I} \sim N(\boldsymbol{\mu}_{\boldsymbol{\eta}}(\boldsymbol{y}_{-I}), \boldsymbol{\Sigma}_{\boldsymbol{\eta}}(\boldsymbol{y}_{-I}))$ and likelihood $\pi_G(\boldsymbol{y}_{I}|\boldsymbol{\eta}_{I}) \propto \exp\bigg\{- \frac{1}{2}\boldsymbol{\eta}_{I}^T\boldsymbol{C}(\boldsymbol{y}_{I})\boldsymbol{\eta}_{I} + \boldsymbol{b}(\boldsymbol{y}_{I})\boldsymbol{\eta}_{I} \bigg\}$. Now, we need to use the posterior and the likelihood to obtain the prior.

If $\boldsymbol{\Sigma}_{\boldsymbol{\eta}_{I}}(\boldsymbol{y})$ is full rank, we have $\boldsymbol{Q}_{\boldsymbol{\eta}_I}(\boldsymbol{y}) = \boldsymbol{\Sigma}_{\boldsymbol{\eta}_{I}}(\boldsymbol{y})^{-1}$ and $\boldsymbol{b}_{\boldsymbol{\eta}_I}(\boldsymbol{y}) = \boldsymbol{Q}_{\boldsymbol{\eta}_{I}}(\boldsymbol{y})\boldsymbol{\mu}_{\boldsymbol{\eta}_{I}}(\boldsymbol{y})$. By conjugacy of Gaussian prior and Gaussian likelihood, $\boldsymbol{Q}_{\boldsymbol{\eta}_I}(\boldsymbol{y}_{-I}) = \boldsymbol{Q}_{\boldsymbol{\eta}_{I}}(\boldsymbol{y}) - \boldsymbol{C}(\boldsymbol{y}_{I})$ and $\boldsymbol{b}_{\boldsymbol{\eta}_I}(\boldsymbol{y}_{-I}) = \boldsymbol{Q}_{\boldsymbol{\eta}_{I}}(\boldsymbol{y})\boldsymbol{\mu}_{\boldsymbol{\eta}_{I}}(\boldsymbol{y}) - \boldsymbol{b}(\boldsymbol{y}_{I})$. Then we have desired $\boldsymbol{\mu}_{\boldsymbol{\eta}_{I}}(\boldsymbol{y}_{-I})$ and $\boldsymbol{\Sigma}_{\boldsymbol{\eta}_{I}}(\boldsymbol{y}_{-I})$.

If $\boldsymbol{\Sigma}_{\boldsymbol{\eta}_{I}}(\boldsymbol{y})$ is singular, we let $\boldsymbol{\eta}|\boldsymbol{y} = \boldsymbol{B} \boldsymbol{z}|\boldsymbol{y}$, where $\boldsymbol{B} = \boldsymbol{V}\boldsymbol{\Lambda}$ with $\boldsymbol{V}$ containing eigenvectors corresponding to non-zero eigenvalues, $\boldsymbol{\Lambda}$ containing square root of non-zero eigenvalues on its diagonal, and $\boldsymbol{z}|\boldsymbol{y} \sim N(\boldsymbol{\mu}_{\boldsymbol{z}}(\boldsymbol{y}),\boldsymbol{\mathcal{I}})$, where $\boldsymbol{\mathcal{I}}$ is an identity matrix and $\boldsymbol{\mu}_{\boldsymbol{z}}(\boldsymbol{y}) = \boldsymbol{B}^T\boldsymbol{\mu}_{\boldsymbol{\eta}_I}(\boldsymbol{y})$. By conjugacy, we have $\boldsymbol{Q}_{\boldsymbol{z}}(\boldsymbol{y}_{-I}) = \boldsymbol{\mathcal{I}} - \boldsymbol{B}^T \boldsymbol{C}(\boldsymbol{y}_{I})\boldsymbol{B}$ and $\boldsymbol{b}_{\boldsymbol{z}}(\boldsymbol{y}_{-I}) = \boldsymbol{\mu}_{\boldsymbol{z}}(\boldsymbol{y}) - \boldsymbol{B}^T \boldsymbol{b}(\boldsymbol{y}_{I})$. It is followed by $\boldsymbol{\mu}_{\boldsymbol{z}}(\boldsymbol{y}_{-I}) = \boldsymbol{Q}_{\boldsymbol{z}}(\boldsymbol{y}_{-I})^{-1}\boldsymbol{b}_{\boldsymbol{z}}(\boldsymbol{y}_{-I})$. Then mean and covariance of $\boldsymbol{z}|\boldsymbol{y}_{-I}$ is $\boldsymbol{\mu}_{\boldsymbol{\eta}}(\boldsymbol{y}_{-I}) = \boldsymbol{B}\boldsymbol{\mu}_{\boldsymbol{z}}(\boldsymbol{y}_{-I})$, $\boldsymbol{\Sigma}_{\boldsymbol{\eta}}(\boldsymbol{y}_{-I}) = \boldsymbol{B}\boldsymbol{\Sigma}_{\boldsymbol{z}}(\boldsymbol{y}_{-I})\boldsymbol{B}^T$.

\section[On the computation of Sigma and mu with Linear Constraints]{On the computation of $\boldsymbol{\Sigma}_{\boldsymbol{\eta}_{I_i}}(\boldsymbol{\theta},\boldsymbol{y})$ and $\boldsymbol{\mu}_{\boldsymbol{\eta}_{I_i}}(\boldsymbol{\theta},\boldsymbol{y})$ with Linear Constraints}
\label{appendix:sig_eta}

We start by illustrating how to compute $\boldsymbol{\Sigma}_{\boldsymbol{\eta}_{I_i}}(\boldsymbol{\theta},\boldsymbol{y})$ and $\boldsymbol{\mu}_{\boldsymbol{\eta}_{I_i}}(\boldsymbol{\theta},\boldsymbol{y})$  without linear constraints. $\boldsymbol{\mu}_{\boldsymbol{\eta}_{I_i}}(\boldsymbol{\theta},\boldsymbol{y})$ is simply obtained by $\boldsymbol{\mu}_{\boldsymbol{\eta}_{I_i}}(\boldsymbol{\theta},\boldsymbol{y}) = \boldsymbol{A}_{I_i}\boldsymbol{\mu}_{\boldsymbol{f}}(\boldsymbol{\theta},\boldsymbol{y})$. However, we never store large dense matrix like $\boldsymbol{Q}_{\boldsymbol{f}}(\boldsymbol{\theta},\boldsymbol{y})^{-1}$. Thus, $\boldsymbol{\Sigma}_{\boldsymbol{\eta}_{I_i}}(\boldsymbol{\theta},\boldsymbol{y})$ cannot be obtained by using matrix multiplication $\boldsymbol{A}_{I_i}\boldsymbol{Q}_{\boldsymbol{f}}(\boldsymbol{\theta},\boldsymbol{y})^{-1}\boldsymbol{A}_{I_i}^T$. Instead, we compute $\boldsymbol{\Sigma}_{\boldsymbol{\eta}}(\boldsymbol{\theta},\boldsymbol{y})$ entry by entry and use the result to fill in entries of $\boldsymbol{\Sigma}_{\boldsymbol{\eta}_{I_i}}(\boldsymbol{\theta},\boldsymbol{y})$. We compute $\boldsymbol{\Sigma}_{\boldsymbol{\eta}}(\boldsymbol{\theta},\boldsymbol{y})_{i,j}$ by solving $$\boldsymbol{Q}_{\boldsymbol{f}}(\boldsymbol{\theta},\boldsymbol{y})\boldsymbol{x} = \boldsymbol{A_i}$$ and $\boldsymbol{\Sigma}_{\boldsymbol{\eta}}(\boldsymbol{\theta},\boldsymbol{y})_{i,j} = \boldsymbol{A_j}\boldsymbol{x}$. The computation is fast because $\boldsymbol{A}$ and $\boldsymbol{Q}_{\boldsymbol{f}}(\boldsymbol{\theta},\boldsymbol{y})$ are sparse, and the factorization of $\boldsymbol{Q}_{\boldsymbol{f}}(\boldsymbol{\theta},\boldsymbol{y})$ is reused. 

When linear constraints $\boldsymbol{\mathcal{C}}\boldsymbol{f} = \boldsymbol{e}$ are applied on $\boldsymbol{f}$, we have 
\begin{equation*}
	\begin{split}
    & \boldsymbol{\Sigma}_{\boldsymbol{f}}(\boldsymbol{\theta},\boldsymbol{y})^* = \boldsymbol{Q}_{\boldsymbol{f}}(\boldsymbol{\theta},\boldsymbol{y})^{-1} - \boldsymbol{Q}_{\boldsymbol{f}}(\boldsymbol{\theta},\boldsymbol{y})^{-1}\boldsymbol{\mathcal{C}}^T(\boldsymbol{\mathcal{C}}\boldsymbol{Q}_{\boldsymbol{f}}(\boldsymbol{\theta},\boldsymbol{y})^{-1}\boldsymbol{\mathcal{C}}^T)^{-1}\boldsymbol{\mathcal{C}}\boldsymbol{Q}_{\boldsymbol{f}}(\boldsymbol{\theta},\boldsymbol{y})^{-1},\\
    & \boldsymbol{\mu}_{\boldsymbol{f}}(\boldsymbol{\theta},\boldsymbol{y})^* = \boldsymbol{\mu}_{\boldsymbol{f}}(\boldsymbol{\theta},\boldsymbol{y}) - \boldsymbol{Q}_{\boldsymbol{f}}(\boldsymbol{\theta},\boldsymbol{y})^{-1}\boldsymbol{\mathcal{C}}^T(\boldsymbol{\mathcal{C}}\boldsymbol{Q}_{\boldsymbol{f}}(\boldsymbol{\theta},\boldsymbol{y})^{-1}\boldsymbol{\mathcal{C}}^T)^{-1}(\boldsymbol{\mathcal{C}} \boldsymbol{\mu}_{\boldsymbol{f}} - \boldsymbol{e}),
    \end{split}
\end{equation*}
where $\boldsymbol{\Sigma}_{\boldsymbol{f}}(\boldsymbol{\theta},\boldsymbol{y})^*$ and $\boldsymbol{\mu}_{\boldsymbol{f}}(\boldsymbol{\theta},\boldsymbol{y})^*$ are the mean and the covariance matrix after applying constraints \cite{rue2005gaussian}. Because $\boldsymbol{\mu}_{\boldsymbol{f}}(\boldsymbol{\theta},\boldsymbol{y})^*$, is always stored, the computation of $\boldsymbol{\mu}_{\boldsymbol{\eta}_{I_i}}(\boldsymbol{\theta},\boldsymbol{y})$ is simple. We need to propagate the effects of linear constraints to $\boldsymbol{\Sigma}_{\boldsymbol{\eta}}(\boldsymbol{\theta},\boldsymbol{y})_{i,j}$.  This is achieved by computing \cite{rue2005gaussian}
$$\boldsymbol{x}^* = \boldsymbol{x} - \boldsymbol{Q}_{\boldsymbol{f}}(\boldsymbol{\theta},\boldsymbol{y})^{-1}\boldsymbol{\mathcal{C}}^T(\boldsymbol{\mathcal{C}}\boldsymbol{Q}_{\boldsymbol{f}}(\boldsymbol{\theta},\boldsymbol{y})^{-1}\boldsymbol{\mathcal{C}}^T)^{-1}\boldsymbol{\mathcal{C}}\boldsymbol{x},$$
where $\boldsymbol{x}$ solves $\boldsymbol{Q}_{\boldsymbol{f}}(\boldsymbol{\theta},\boldsymbol{y})\boldsymbol{x} = \boldsymbol{A_i}$. Then 
$\boldsymbol{\Sigma}_{\boldsymbol{\eta}}(\boldsymbol{\theta},\boldsymbol{y})^*_{i,j} = \boldsymbol{A_j}\boldsymbol{x}^*.$


\bibliography{references}

\begin{thebibliography}{10}

\bibitem{stone1974cross}
Mervyn Stone.
\newblock Cross-validatory choice and assessment of statistical predictions.
\newblock {\em Journal of the royal statistical society: Series B (Methodological)}, 36(2):111--133, 1974.

\bibitem{Geisser1979}
Seymour Geisser and William~F. Eddy.
\newblock A predictive approach to model selection.
\newblock {\em Journal of the American Statistical Association}, 74(365):153--160, 1979.

\bibitem{gneiting2007strictly}
Tilmann Gneiting and Adrian~E Raftery.
\newblock Strictly proper scoring rules, prediction, and estimation.
\newblock {\em Journal of the American statistical Association}, 102(477):359--378, 2007.

\bibitem{vehtari2012survey}
Aki Vehtari, Janne Ojanen, et~al.
\newblock A survey of bayesian predictive methods for model assessment, selection and comparison.
\newblock {\em Statistics Surveys}, 6:142--228, 2012.

\bibitem{burman1994cross}
Prabir Burman, Edmond Chow, and Deborah Nolan.
\newblock A cross-validatory method for dependent data.
\newblock {\em Biometrika}, 81(2):351--358, 1994.

\bibitem{mcquarrie1998regression}
Allan~DR McQuarrie and Chih-Ling Tsai.
\newblock {\em Regression and time series model selection}.
\newblock World Scientific, 1998.

\bibitem{bergmeir2012use}
Christoph Bergmeir and Jos{\'e}~M Ben{\'\i}tez.
\newblock On the use of cross-validation for time series predictor evaluation.
\newblock {\em Information Sciences}, 191:192--213, 2012.

\bibitem{bergmeir2018note}
Christoph Bergmeir, Rob~J Hyndman, and Bonsoo Koo.
\newblock A note on the validity of cross-validation for evaluating autoregressive time series prediction.
\newblock {\em Computational Statistics \& Data Analysis}, 120:70--83, 2018.

\bibitem{burkner2020approximate}
Paul-Christian B{\"u}rkner, Jonah Gabry, and Aki Vehtari.
\newblock Approximate leave-future-out cross-validation for bayesian time series models.
\newblock {\em Journal of Statistical Computation and Simulation}, 90(14):2499--2523, 2020.

\bibitem{cerqueira2020evaluating}
Vitor Cerqueira, Luis Torgo, and Igor Mozeti{\v{c}}.
\newblock Evaluating time series forecasting models: An empirical study on performance estimation methods.
\newblock {\em Machine Learning}, 109(11):1997--2028, 2020.

\bibitem{valavi2018blockcv}
Roozbeh Valavi, Jane Elith, Jos{\'e}~J Lahoz-Monfort, and Gurutzeta Guillera-Arroita.
\newblock blockcv: An r package for generating spatially or environmentally separated folds for k-fold cross-validation of species distribution models.
\newblock {\em Biorxiv}, page 357798, 2018.

\bibitem{hartigan1979k}
John~A Hartigan, Manchek~A Wong, et~al.
\newblock A k-means clustering algorithm.
\newblock {\em Applied statistics}, 28(1):100--108, 1979.

\bibitem{saeb2017need}
Sohrab Saeb, Luca Lonini, Arun Jayaraman, David~C Mohr, and Konrad~P Kording.
\newblock The need to approximate the use-case in clinical machine learning.
\newblock {\em Gigascience}, 6(5):gix019, 2017.

\bibitem{gelman1995bayesian}
Andrew Gelman, John~B Carlin, Hal~S Stern, and Donald~B Rubin.
\newblock {\em Bayesian data analysis}.
\newblock Chapman and Hall/CRC, 1995.

\bibitem{racine2000consistent}
Jeff Racine.
\newblock Consistent cross-validatory model-selection for dependent data: hv-block cross-validation.
\newblock {\em Journal of econometrics}, 99(1):39--61, 2000.

\bibitem{merkle2019bayesian}
Edgar~C Merkle, Daniel Furr, and Sophia Rabe-Hesketh.
\newblock Bayesian comparison of latent variable models: Conditional versus marginal likelihoods.
\newblock {\em Psychometrika}, 84:802--829, 2019.

\bibitem{roberts2017cross}
David~R Roberts, Volker Bahn, Simone Ciuti, Mark~S Boyce, Jane Elith, Gurutzeta Guillera-Arroita, Severin Hauenstein, Jos{\'e}~J Lahoz-Monfort, Boris Schr{\"o}der, Wilfried Thuiller, et~al.
\newblock Cross-validation strategies for data with temporal, spatial, hierarchical, or phylogenetic structure.
\newblock {\em Ecography}, 40(8):913--929, 2017.

\bibitem{rabinowicz2022cross}
Assaf Rabinowicz and Saharon Rosset.
\newblock Cross-validation for correlated data.
\newblock {\em Journal of the American Statistical Association}, 117(538):718--731, 2022.

\bibitem{adin2024automatic}
Aritz Adin, Elias~Teixeira Krainski, Amanda Lenzi, Zhedong Liu, Joaqu{\'\i}n Mart{\'\i}nez-Minaya, and Haavard Rue.
\newblock Automatic cross-validation in structured models: Is it time to leave out leave-one-out?
\newblock {\em Spatial Statistics}, page 100843, 2024.

\bibitem{rue2009approximate}
H{\aa}vard Rue, Sara Martino, and Nicolas Chopin.
\newblock Approximate bayesian inference for latent gaussian models by using integrated nested laplace approximations.
\newblock {\em Journal of the royal statistical society: Series b (statistical methodology)}, 71(2):319--392, 2009.

\bibitem{rue2017bayesian}
H{\aa}vard Rue, Andrea Riebler, Sigrunn~H S{\o}rbye, Janine~B Illian, Daniel~P Simpson, and Finn~K Lindgren.
\newblock Bayesian computing with inla: a review.
\newblock {\em Annual Review of Statistics and Its Application}, 4:395--421, 2017.

\bibitem{van2021correcting}
Janet van Niekerk and Haavard Rue.
\newblock Low-rank variational bayes correction to the laplace method.
\newblock {\em Journal of Machine Learning Research}, 25(62):1--25, 2024.

\bibitem{van2023new}
Janet Van~Niekerk, Elias Krainski, Denis Rustand, and H{\aa}vard Rue.
\newblock A new avenue for bayesian inference with inla.
\newblock {\em Computational Statistics \& Data Analysis}, 181:107692, 2023.

\bibitem{akaike1973information}
H~Akaike.
\newblock Information theory and an extension of the maximum likelihood principle.
\newblock In {\em 2nd International Symposium on Information Theory}, pages 267--281. Akad{\'e}miai Kiad{\'o} Location Budapest, Hungary, 1973.

\bibitem{watanabe2010asymptotic}
Sumio Watanabe.
\newblock Asymptotic equivalence of bayes cross validation and widely applicable information criterion in singular learning theory.
\newblock {\em Journal of machine learning research}, 11(12), 2010.

\bibitem{stone1977asymptotic}
Mervyn Stone.
\newblock An asymptotic equivalence of choice of model by cross-validation and akaike's criterion.
\newblock {\em Journal of the Royal Statistical Society: Series B (Methodological)}, 39(1):44--47, 1977.

\bibitem{yang2007consistency}
Yuhong Yang.
\newblock Consistency of cross validation for comparing regression procedures.
\newblock {\em The Annals of Statistics}, 35(6):2450--2473, 2007.

\bibitem{shao1993linear}
Jun Shao.
\newblock Linear model selection by cross-validation.
\newblock {\em Journal of the American statistical Association}, 88(422):486--494, 1993.

\bibitem{wang2018bayesian}
Xiaofeng Wang, Yu~Ryan Yue, and Julian~J Faraway.
\newblock {\em Bayesian regression modeling with INLA}.
\newblock CRC Press, 2018.

\bibitem{krainski2018advanced}
Elias Krainski, Virgilio G{\'o}mez-Rubio, Haakon Bakka, Amanda Lenzi, Daniela Castro-Camilo, Daniel Simpson, Finn Lindgren, and H{\aa}vard Rue.
\newblock {\em Advanced spatial modeling with stochastic partial differential equations using R and INLA}.
\newblock Chapman and Hall/CRC, 2018.

\bibitem{gomez2020bayesian}
Virgilio G{\'o}mez-Rubio.
\newblock {\em Bayesian inference with INLA}.
\newblock CRC Press, 2020.

\bibitem{rue2005gaussian}
Havard Rue and Leonhard Held.
\newblock {\em Gaussian Markov random fields: theory and applications}.
\newblock Chapman and Hall/CRC, 2005.

\bibitem{liu1994note}
Qing Liu and Donald~A Pierce.
\newblock A note on gauss—hermite quadrature.
\newblock {\em Biometrika}, 81(3):624--629, 1994.

\bibitem{lowe2021combined}
Rachel Lowe, Sophie~A Lee, Kathleen~M O'Reilly, Oliver~J Brady, Leonardo Bastos, Gabriel Carrasco-Escobar, Rafael de~Castro~Cat{\~a}o, Felipe~J Col{\'o}n-Gonz{\'a}lez, Christovam Barcellos, Marilia~S{\'a} Carvalho, et~al.
\newblock Combined effects of hydrometeorological hazards and urbanisation on dengue risk in brazil: a spatiotemporal modelling study.
\newblock {\em The Lancet Planetary Health}, 5(4):e209--e219, 2021.

\bibitem{Rstan}
{Stan Development Team}.
\newblock {RStan}: the {R} interface to {Stan}, 2022.
\newblock R package version 2.21.5.

\bibitem{besag1991bayesian}
Julian Besag, Jeremy York, and Annie Molli{\'e}.
\newblock Bayesian image restoration, with two applications in spatial statistics.
\newblock {\em Annals of the institute of statistical mathematics}, 43(1):1--20, 1991.

\bibitem{wakefield2000bayesian}
Jonathan~C Wakefield, NG~Best, and L~Waller.
\newblock Bayesian approaches to disease mapping.
\newblock {\em Spatial epidemiology: methods and applications}, pages 104--127, 2000.

\bibitem{held2005towards}
Leonhard Held, Isabel Nat{\'a}rio, Sarah~Elaine Fenton, H{\aa}vard Rue, and Nikolaus Becker.
\newblock Towards joint disease mapping.
\newblock {\em Statistical methods in medical research}, 14(1):61--82, 2005.

\bibitem{rachel_lowe_2021_4632205}
Rachel Lowe.
\newblock {Data and R code to accompany 'Combined effects of hydrometeorological hazards and urbanisation on dengue risk in Brazil: a spatiotemporal modelling study'}, March 2021.
\newblock Software, version v1.0.0.

\bibitem{riebler2016intuitive}
Andrea Riebler, Sigrunn~H S{\o}rbye, Daniel Simpson, and H{\aa}vard Rue.
\newblock An intuitive bayesian spatial model for disease mapping that accounts for scaling.
\newblock {\em Statistical methods in medical research}, 25(4):1145--1165, 2016.

\bibitem{vehtari2019limitations}
Aki Vehtari, Daniel~P Simpson, Yuling Yao, and Andrew Gelman.
\newblock Limitations of “limitations of bayesian leave-one-out cross-validation for model selection”.
\newblock {\em Computational Brain \& Behavior}, 2(1):22--27, 2019.

\end{thebibliography}
\bibliographystyle{unsrt}
\end{document}